\documentclass[letterpaper, unpublished, onecolumn, accepted=2024-11-14]{quantumarticle}
\pdfoutput=1

\usepackage{graphicx, amsmath, amssymb}
\usepackage{physics}
\usepackage{xcolor}
\usepackage[numbers]{natbib}

\usepackage{hyperref}
\hypersetup{
    colorlinks=true,
    linkcolor=blue,
    filecolor=magenta,      
    urlcolor=cyan,
    pdftitle={Overleaf Example},
    pdfpagemode=FullScreen,
    }

\title{
    Bosonic Pauli+: Efficient Simulation of Concatenated Gottesman--Kitaev--Preskill Codes
}
\author[1]{Florian Hopfmueller}
\email{florian@nordquantique.ca}
\author[1]{Maxime Tremblay}
\author[1]{Philippe St--Jean}
\author[2]{Baptiste Royer}
\author[1]{Marc--Antoine Lemonde}
\email{marc-antoine@nordquantique.ca}
\affil[1]{Nord Quantique, Sherbrooke, Qu\'ebec, Canada}
\affil[2]{Institut Quantique and D\'epartment de Physique, Universit\'e de Sherbrooke, Qu\'ebec, Canada}
\date{Oct 16, 2024}

\newcommand{\hilbert}{\mathcal{H}}
\newcommand{\suchthat}{\,:\,}

\newcommand{\channel}{\mathcal{C}}
\newcommand{\ptmp}{{\channel_{\text{PTM}+}}}
\newcommand{\bpp}{{\channel_{\text{BP}+}}}
\newcommand{\pauli}{{\hat{\sigma}}}
\newcommand{\sBs}{\text{sBs}}

\newcommand{\PR}{{\mathcal P_{\mathcal R}}}

\usepackage{newfloat}
\DeclareFloatingEnvironment[
    fileext=los,
    listname={List of Algorithms},
    name=Algorithm,
    placement=tbhp,
]{algorithm}

\begin{document}

\maketitle

\begin{abstract}
    A promising route towards fault--tolerant quantum error correction is the concatenation of a Gottesman--Kitaev--Preskill (GKP) code with a qubit code. Development of such concatenated codes requires simulation tools which realistically model noise, while being able to simulate the dynamics of many modes. However, so far, large--scale simulation tools for concatenated GKP codes have been limited to idealized noise models and GKP code implementations. Here, we introduce the Bosonic Pauli+ model (BP+), which can be simulated efficiently for a large number of modes, while capturing the rich dynamics in the bosonic multi--mode Hilbert space. We demonstrate the method by simulating a hybrid surface code, where the data qubits are finite--energy GKP qubits stabilized using the small--Big--small (sBs) protocol, and the syndrome qubits are standard two--level systems. Using BP+, we present logical error rates of such an implementation. Confidence in the accuracy of the method is gained by comparing its predictions with full time evolution simulations for several relevant quantum circuits. While developed specifically for GKP qubits stabilized using the sBs protocol, the mathematical structure of BP+ is generic and may be applicable also to the simulation of concatenations using other bosonic codes.
\end{abstract}

\section{Introduction}
A promising proposal to reach fault--tolerant quantum computation is to exploit the large Hilbert space of a quantum harmonic oscillator to redundantly encode logical information, thus implementing a first layer of error correction at the single--mode level using a bosonic quantum error correcting (QEC) code. Then any remaining errors may be dealt with by concatenating these error--corrected bosonic modes with an outer qubit QEC code. 
A promising single--mode code to implement this vision is the Gottesman--Kitaev--Preskill (GKP) code, which 
has first been suggested in ref.~\cite{gottesman_encoding_2001}, 
and has recently seen much attention and success in experimental implementation \cite{fluhmann_encoding_2019,  de_neeve_error_2022, campagne-ibarcq_quantum_2020, sivak_real-time_2023, konno_propagating_2023, lachance-quirion_autonomous_2023, kudra_robust_2022, matsos_robust_2024}.
The approach of concatenating a GKP code and a qubit outer code could greatly reduce the hardware overhead 
associated with QEC, compared to the standard approach of a qubit code built from unprotected two--level systems (TLS), as was recently argued by many authors \cite{
    zhang_concatenation_2023,
    li_correcting_2023,
    noh_fault-tolerant_2020,
    raveendran_finite_2022,
    noh_low-overhead_2022,
    zhang_quantum_2021,
    vuillot_quantum_2019,
    xu_qubit-oscillator_2023}.

However, many questions are open within this general paradigm.
For example, even leaving aside the implementation of the GKP code itself, it is not clear which qubit outer code should be used,
nor is it known whether the outer code should best be implemented with all GKP qubits, or using mixed qubit species.
Additionally, it is not understood in this context how to best measure the outer code stabilizer generators,
or whether multi--mode GKP codes \cite{gottesman_encoding_2001, harrington_achievable_2001, royer_encoding_2022, conrad_gottesman-kitaev-preskill_2022, lin_closest_2023} should be employed, which blur the distinction between inner and outer codes.
Furthermore, it is not understood which hardware coherence times and gate fidelities need to be achieved in order to implement large--scale quantum computation in this approach, nor how the GKP code compares to other bosonic codes as a building block of a concatenated code, such as dual--rail qubits \cite{levine_demonstrating_2023, chou_demonstrating_2023} or cat qubits \cite{chamberland_building_2022, guillaud_repetition_2019}.

\begin{figure}
	\includegraphics{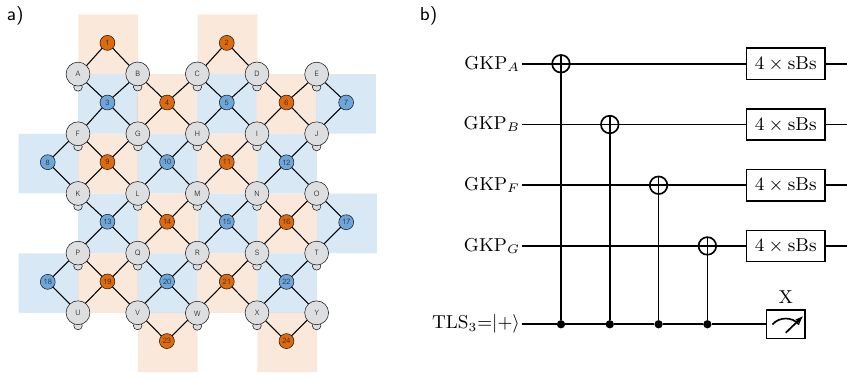}
 
 %    \begin{minipage}{.45\textwidth}
 %    \includegraphics[width=.9\textwidth]{surface code illustration_v2_231101.pdf}
 %    \end{minipage}
 %    \begin{minipage}{.45\textwidth}{\small
	% \begin{quantikz}
	% 	\lstick{GKP$_A$} & \targ{} & \qw & \qw & \qw & \gate{4 \times \sBs} & \qw\\
	% 	\lstick{GKP$_B$} & \qw & \targ{} & \qw & \qw & \gate{4 \times \sBs} & \qw\\
	% 	\lstick{GKP$_F$} & \qw & \qw & \targ{} & \qw & \gate{4 \times \sBs} & \qw\\
	% 	\lstick{GKP$_G$} & \qw & \qw & \qw & \targ{} & \gate{4 \times \sBs} & \qw\\
	% 	\lstick{TLS$_3$=$\ket{+}$}  & \ctrl{-4} & \ctrl{-3} & \ctrl{-2} & \ctrl{-1} & \meter{X} &
	% \end{quantikz}}
 %    \end{minipage}
    \caption{Illustrations of the concatenated code simulated in this work. a): The outer code is a $d=5$ rotated surface code, with data qubits (large gray circles, labelled A--Y) which are GKP modes, and syndrome qubits (smaller blue and red circles, labelled 1--24) which are two--level systems and are used to measure the outer--code stabilizers. The required connectivity between modes is shown with black lines.
    In addition, each GKP mode has an auxiliary two--level system (TLS) used to stabilize the mode using the $\sBs$ protocol. The auxiliary TLS is illustrated with small gray circles, and the $\sBs$ protocol is shown in fig.~\ref{fig:sbs_circuit}.
    b) Outer code parity check circuit, reading out the outer--code stabilizer $\hat X_A \hat X_B \hat X_F \hat X_G$. After each CNOT gate, the participating GKP mode is stabilized by applying four rounds of the $\sBs$ protocol.}
    \label{fig:surface_code_illustration}
\end{figure}

A prerequisite to answer these questions is the ability to perform numerical simulations. 
These simulations must be able to scale efficiently to a large number of GKP modes, 
while remaining sufficiently accurate to make reliable predictions about the performance that could be expected in a hardware implementation.
This need parallels the situation for qubit codes built from standard TLS, 
where the fast pace of innovation in recent research would have been unthinkable without the 
availability of fast Clifford simulations \cite{gottesman_heisenberg_1998, aaronson_improved_2004} with experimentally informed noise models.

To date, such simulation tools are lacking for concatenated GKP codes.
Straightforward application of standard techniques falls short of the goal: 
Classically simulating the full quantum dynamics of the multi--mode system requires exorbitant resources even for a few modes.
Applying standard Clifford simulations, 
which can efficiently simulate qubit codes built from standard TLS, 
fails to accurately model the system:
Bosonic qubits are more than just TLS, but they have additional non--logical degrees of freedom.
These degrees of freedom can lead to time--correlated errors acting on the logical information, and can also be measured, 
providing valuable inner code syndrome information --- features which are not captured by standard Clifford simulations.

An existing simulation method for the concatenation of a GKP code with a qubit code 
\cite{noh_fault-tolerant_2020, noh_low-overhead_2022} 
works with simple analytical models for the code implementation and errors.
There, errors are modelled as displacement errors from a Gaussian distribution acting on infinite--energy GKP states, 
and the implementation of GKP error correction involves the use of auxiliary GKP qubits and homodyne measurements.
This method gives insight into theoretical performance limits and propagation of errors,
but does not faithfully model the implementation and physical noise channels of recent superconducting hardware implementations  {where stabilization is implemented using an auxiliary TLS}
\cite{sivak_real-time_2023, lachance-quirion_autonomous_2023}.

Here, we introduce the Bosonic Pauli+ (BP+) model and simulation method.
BP+ enriches standard Clifford simulations in a way that captures the dynamics of the multi--mode bosonic system, 
while still being efficiently simulatable.
The model relies on choosing a basis of bosonic Hilbert space, 
and we introduce the $\sBs$ basis, which is designed for the simulation of finite--energy GKP qubits stabilized using the small--Big--small ($\sBs$) protocol \cite{royer_stabilization_2020, de_neeve_error_2022}.
We demonstrate the model, and give numerical evidence of its accuracy, 
for a particular concatenation of a single--mode GKP code and a qubit surface code.
The key ideas are general, and may be applicable also to other implementations of the concatenation, 
and to concatenations of other GKP implementations and other bosonic codes.

The BP+ method draws inspiration from two lines of work:
Firstly, even standard qubits, like transmons, are often not just TLS in practice, 
but have additional non--computational leakage levels. 
The presence of leakage levels can cause correlated noise to act on the computational levels which cannot be explained by standard Clifford simulations.
In order to more accurately model the dynamics of a QEC code in the presence of leakage levels, standard Clifford simulations have been extended in the Pauli+ method \cite{google_quantum_ai_suppressing_2023, marshall_incoherent_2023}, building on earlier work \cite{fowler_coping_2013}.
In a Pauli+ simulation, there is an additional discrete variable for each physical qubit, specifying whether the qubit presently occupies a computational level or a leakage level.
The probabilistic dynamics of the additional variables are governed by transition rates into, from, and among the leakage levels, and the dynamics of the computational quantum state may depend on the additional variables. We adopt these ideas in BP+. Related ideas have also been used in a simulation of a repetition code using bosonic cat qubits \cite{regent_high-performance_2023}.
The Pauli+ method has shown success in explaining data obtained experimentally from an implementation of the $d=3$ and $d=5$ surface codes \cite{google_quantum_ai_suppressing_2023}. One of the approximations entering the Pauli+ method was recently formalized and analyzed more closely in ref.~\cite{marshall_incoherent_2023}.

The second line of work that inspired the BP+ method is the effort to understand the dynamics of GKP qubits in the presence of errors
by decomposing the Hilbert space of the GKP mode as the tensor product a two--level logical subsystem, and an error subsystem accounting for the remaining degrees of freedom.
Subsystem decompositions well--suited to infinite--energy GKP qubits include the Zak basis \cite{zak_finite_1967, pantaleoni_zak_2023} and
the closely related GKP stabilizer subsystem decomposition \cite{shaw_stabilizer_2022}, both of which feature continuous error degrees of freedom.
For the more realistic finite--energy GKP qubits which are considered here, 
a decomposition with discrete error degrees of freedom is proposed in ref.~\cite{sivak_real-time_2023}, 
which gives some insight into the dynamics of finite--energy GKP qubits stabilized using the $\sBs$ protocol.

We apply the BP+ method to simulate a rotated surface code \cite{horsman_surface_2012, fowler_surface_2012}. Each data qubit is a GKP qubit, and the qubits used to read out the stabilizers of the surface code are TLS. This makes the concatenated code a hybrid code in the terminology of ref.~\cite{grimsmo_quantum_2021}.
The high--level structure of this concatenation is illustrated in fig.~\ref{fig:surface_code_illustration}.
We choose a physical noise model which captures the effects of decay and dephasing of the participating modes during the stabilization of the GKP code
and the gates implementing the surface code.
BP+ models are extracted from time evolution simulations using the physical noise model.

The structure of the rest of this paper is as follows.
Section \ref{sec:bos_pauli_plus} presents the  mathematical structure of the BP+ model and simulation method.
Section \ref{sec:sbs_basis} presents the $\sBs$ basis, along with reminders about finite--energy GKP qubits and the $\sBs$ protocol.
Section \ref{sec:gates} presents details of the physical noise model, and extracts BP+ models for $\sBs$ stabilization and the CNOT gates required for the surface code implementation.
Section \ref{sec:gaining_confidence} gathers empirical evidence for the accuracy of the BP+ approximation, applied to the concatenation considered here.
Section \ref{sec:surface_code_sims} presents logical error rates of the concatenated code for three different outer--code decoders.
Section \ref{sec:conclusion} concludes the paper, and appendices contain technical details and additional simulation results.

\begin{figure}
    \centering
    \includegraphics[width=\textwidth]{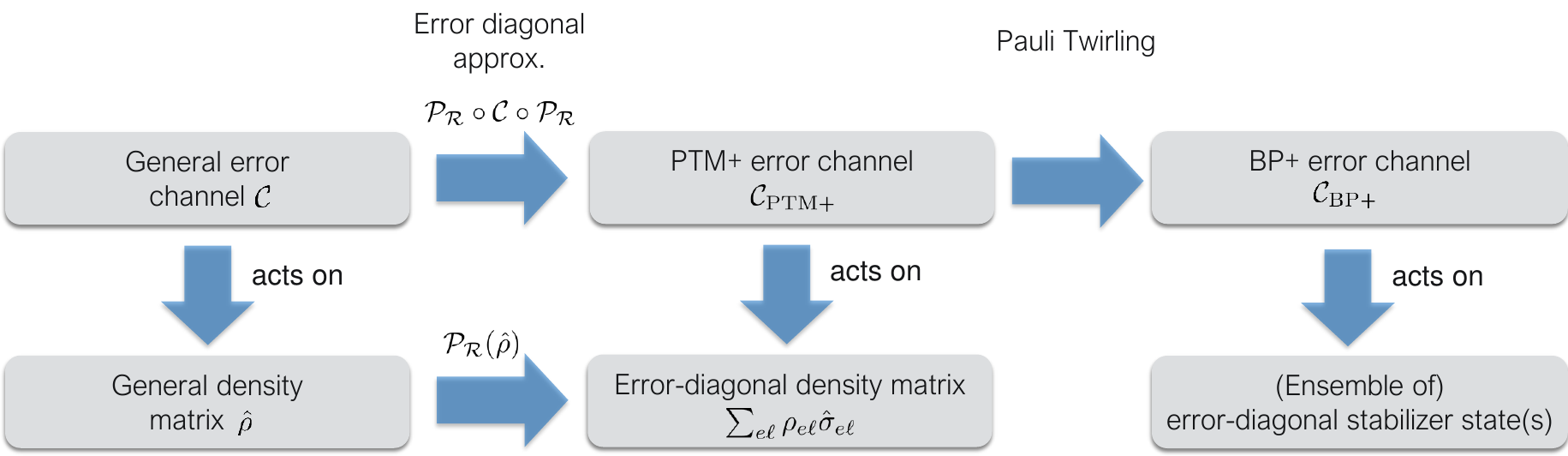}
    \caption{Illustrating the different sets of states, and the channels that act on them, in this work.}
    \label{fig:flowchart}
\end{figure}

\section{The Bosonic Pauli+ Model}
\label{sec:bos_pauli_plus}

The central idea of BP+ is to express the Hilbert space of a bosonic mode as a 
tensor product of a logical and an error Hilbert space, and 
approximate the error Hilbert space as a classical space. Thus, only classical 
mixtures of different basis states of error Hilbert space are allowed.
We call this restricted set of states error--diagonal density matrices, 
and channels acting on this set are termed Pauli Transfer Matrix+ (PTM+) channels.
In a second approximation, the action of a channel on the logical Hilbert 
space is approximated as a Pauli channel, resulting in a BP+ channel.
A BP+ channel can then be parameterized by a transition matrix, 
which indicates the rate of population transfer between the basis states of error Hilbert space,
and a set of Pauli error rates, which specify the Pauli error channel to 
be applied on the logical Hilbert space alongside each error space transition.
This parameterization allows for efficient simulation of noisy Clifford circuits 
containing a large number of modes. The different classes of channels and the states they act on are illustrated in fig.~\ref{fig:flowchart}.

\subsection{Subspace decomposition and error--diagonal states}\label{sec:subspace_decomp_theory}

We consider a bosonic qubit with Hilbert space $\hilbert_B$, on which a cutoff is 
introduced such that $\hilbert_B$ is of finite, even dimension $|\hilbert_B|$.
For our numerical simulations, we will work with a Fock cutoff chosen as an even number large enough to not affect predictions.
The Hilbert space of each qubit is decomposed into two subsystems: a two--dimensional Hilbert space $\hilbert_{L}$
carrying the logical information, 
and an error Hilbert space $\hilbert_{E}$ carrying 
the remaining degrees of freedom as
\begin{equation}
    \hilbert_B = \hilbert_E \otimes \hilbert_L.
\end{equation}
Such a decomposition may be specified in practice by providing an orthonormal basis $B = \qty{\ket{e, \mu}}$ of 
$\hilbert_B$, whose members are indexed by $0 \le e < |\hilbert_B|/2$ and $\mu \in \{0, 1\}$, and then identifying $\ket{e, \mu} = \ket e \otimes \ket{\mu}$.
The basis can thus be written as
\begin{equation}
    B = \qty{ 
        \ket{e} \otimes \ket{\mu}
        \suchthat
        \ket{e} \in \hilbert_E,
        \ket{\mu} \in \hilbert_L
    }.
\end{equation}
Given several quantum modes, each with a subsystem decomposition $\hilbert_{B, n} = 
\hilbert_{E, n} \otimes \hilbert_{L, n}$, a subsystem decomposition of the joint Hilbert 
space is obtained by setting $\hilbert_E = \otimes_n \hilbert_{E, n}$ and 
$\hilbert_L = \otimes_n \hilbert_{L, n}$. If the system under consideration involves TLS 
in addition to bosonic modes, for uniformity of notation we will consider the TLS to have a trivial one--dimensional error Hilbert space.

\begin{figure*}[t]
    \centering
    \begin{tikzpicture}
    \draw (0, 0) node[inner sep=0] {
    \includegraphics[width=.35\textwidth]{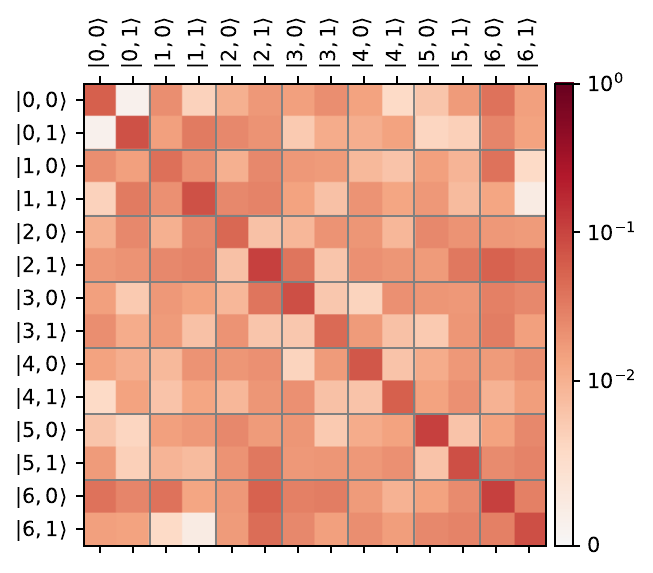}
    \begin{minipage}{.25\textwidth}
    \centering
    \vspace{-4.5cm}
    $\PR$\\
    \vspace{-.5cm}
    \Huge $\rightarrow$
    \end{minipage}
    \includegraphics[width=.35\textwidth]{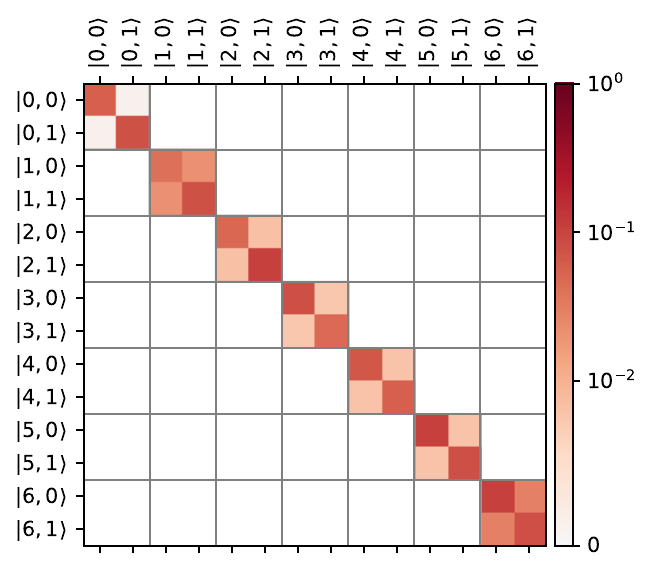}
    };
    \draw (-7.5, 2) node {\small \textsf{a)}};
    \draw (2, 2) node {\small \textsf{b)}};

    \end{tikzpicture}
    \caption{Illustration of the space of error--diagonal density matrices $\mathcal R$ which the PTM+ and BP+ models act on.
    a) Heatmap of the absolute values of the matrix elements of a general random single--mode density matrix $\hat \rho$, in a basis $B = \{ \ket{e, \mu} \}$, where $e$ indexes the error subspace and $\mu$ the logical subspace. b) An error diagonal density matrix obtained as $\mathcal P_{\mathcal R} (\hat \rho) \in \mathcal R$. The coherences between different error sectors $e \neq e'$, which correspond to the block--off--diagonal elements of the shown matrix, are zeroed. The density matrix can be written as $\sum_{e} \op{e} \otimes \hat \rho^{(L)}_e$, with an independent $2\times 2$ logical density matrix $\hat \rho^{(L)}_e$ for each error sector $e$.}
    \label{fig:density_matrix_from_R}
\end{figure*}

To make simulations of large systems tractable,
we restrict $\hilbert_E$ to be a classical space as illustrated in fig.~\ref{fig:density_matrix_from_R}. 
More precisely, we focus on density matrices which are diagonal in the chosen basis of $\hilbert_E$, and are thus contained in the set
\begin{equation}
    \mathcal{R} = 
    \qty{
        \hat \rho \in \mathcal L(\hilbert_B)
        \suchthat 
        \hat \rho = \sum_{e} \op{e} \otimes \hat \rho_e^{(L)},
        \hat \rho_e^{(L)} \in \mathcal L(\hilbert_L)
    },
\end{equation}
 {where $\mathcal L(\hilbert_L)$ denotes the space of linear operators on $\hilbert_L$.}
We refer to such density matrices as error--diagonal density matrices, and refer to the space $\{\ket {e=\tilde e} \otimes \ket{\psi_L}, \ket{\psi_L}\in \hilbert_L\}$ as the error sector $\tilde e$.
The density matrices in $\mathcal{R}$ may feature statistical mixtures between states
in different error sector $e$, with an arbitrary population in each sector given by $\tr \hat\rho_e^{(L)}$.
Also classical correlations between the error Hilbert space and logical Hilbert space can be expressed, 
since there is an independent logical density matrix $\hat \rho_e^{(L)}$ for every error sector $e$. 
Quantum superpositions, or coherences, between error sectors $e \neq e'$ cannot be expressed, 
nor can quantum entanglement between the error and logical Hilbert spaces, or between the error Hilbert spaces of different modes.
The approximation of working with error--diagonal density matrices can be motivated by observing 
that dynamics during the $\sBs$ error--correction protocol are largely independent of coherences between error sectors, 
as we will explore in more detail in section \ref{sec:sbs_basis}.

Any pure states in $\mathcal{R}$ must have a definite error sector index, and can thus be 
described more efficiently by a tuple $\left(\ket{\psi_L} \in \hilbert_L, e \in 
\{0, ..., |\hilbert_E| -1 \}\right).$ 
Here $e$ is a classical integer index, labelling a basis vector of error Hilbert space.
For mixed states, specifying an error--diagonal density matrix of $N$ modes requires $|\hilbert_E| \cdot 4^N/2$ independent complex coefficients, 
in contrast to a general mixed state which requires $|\hilbert_E|^2 \cdot 4^N/2$ complex coefficients.

Different choices of the basis $B$ lead to different spaces $\mathcal{R}$. 
In particular, the choice of basis of $\hilbert_E$ is important, since the distinction between populations (which are modelled in $\mathcal{R}$) and coherences 
(which are not) is basis dependent.
It is therefore important to find a basis $B$ for which restricting to $\mathcal{R}$ is a good 
approximation, in the particular context of a BP+ simulation. We will do so in section \ref{sec:sbs_basis}.

An arbitrary density matrix may be projected onto $\mathcal{R}$ by the quantum channel
$\mathcal P_\mathcal{R}: \mathcal L(\hilbert_B) \rightarrow \mathcal{R}$
acting as
\begin{align}\label{eq:P_R}
    \mathcal{P}_\mathcal{R} (\hat \rho) =
    \sum_e  ( \op{e} \otimes \hat I_L) \cdot \hat \rho \cdot (\op{e} \otimes \hat I_L),
\end{align}
where $\hat I_L$ is the identity on $\hilbert_L$. The channel $\mathcal P_\mathcal{R}$ can be understood as a measurement of the error sector index $e$, where the result of the measurement is discarded.
It decoheres any superpositions between different error space basis vectors $\ket e, \ket {e'}$. 
The channel is a valid completely positive, trace--preserving (CPTP) quantum channel, 
as evidenced by the existence of the Kraus decomposition eq.~\eqref{eq:P_R}~\cite{nielsen_quantum_2010}.
It satisfies $\mathcal P_\mathcal{R} \circ \mathcal P_\mathcal{R} = \mathcal P_\mathcal{R}$.
The effect of $\mathcal P_\mathcal{R}$ is illustrated in fig.~\ref{fig:density_matrix_from_R}.
 {We emphasize that $\mathcal{P}_\mathcal{R}$ will be used to mathematically define the BP+ approximation, 
and it is not necessary to implement it physically.}

\subsection{Pauli Transfer Matrix+ channels}\label{sec:ptmp_channels}
An error--diagonal density matrix of $N$ modes may be conveniently written in the basis
\begin{equation}\label{eq:pauli_error_basis}
    \hat \rho = \sum_{e \ell} \rho_{e\ell} \hat \sigma_{e\ell}, \text{where } \hat \sigma_{e\ell} = \op{e}{e} \otimes \pauli_\ell.
\end{equation}
Here $\ell \in \{I, X, Y, Z\}^N$ labels the Pauli operators on the logical $N$--mode Hilbert space $\hilbert_L$.
This basis satisfies the relation $\tr(\pauli_{e\ell} \pauli_{e' \ell'}) = 2^N \delta_{ee'} \delta_{\ell\ell'}$.
The coefficients $\rho_{e,I}$ 
captures how much population is in an error sector $e$, 
while the remaining coefficients express the logical state, 
which may be different in each sector. Here and elsewhere, we use $I$ as shorthand for $I^{\otimes N}$.

A PTM+ channel $\ptmp: \mathcal{R} \to \mathcal{R}$ is specified
by coefficients $S_{e e' \ell \ell'}$ such that
\begin{align}\label{eq:ptmplus_action}
    \ptmp(\hat \rho)
    = \sum_{e,e',\ell,\ell'} S_{e e' \ell \ell'} \rho_{e' \ell'} \pauli_{e, \ell}.
\end{align}
The PTM+ representation is a generalization of the standard Pauli transfer matrix representation of a multi--qubit channel, which is also referred to as the Louisville representation or as the
superoperator representation in the Pauli basis~\cite{wood_tensor_2015}.
Any channel $\channel$ on $\hilbert_B$ can be approximated as a PTM+ channel
by setting
\begin{equation}\label{eq:ptm_via_projector}
    \ptmp = \PR \circ \channel \circ \PR.
\end{equation}
This approximation is itself a valid quantum channel: 
If $\channel$ is CPTP, then so is $\ptmp$, since also $\PR$ is CPTP.
The coefficients of the PTM+ channel can be extracted from a numerical implementation of $\channel$ by computing
\begin{equation}
    \label{eq:coeff_ptm_trace}
    S_{e e'\ell\ell'} 
    = 
    \frac{1}{2^N} \tr \qty {\pauli_{e \ell} \channel(\pauli_{e' \ell'}) }.
\end{equation}
Computing the coefficients $S_{e e'\ell\ell'}$, for $N$ modes whose joint Hilbert space has been decomposed as $\hilbert_B = \hilbert_E \otimes \hilbert_L$, requires computing $\channel(\pauli_{e' \ell'})$ for 
$4^N \cdot |\hilbert_E|$ many Pauli operators $\pauli_{e' \ell'}$. This is in contrast to full channel tomography, which would require propagating $4^N \cdot |\hilbert_E|^2$ many operators. 
For Hilbert space sizes of order $100$, in the case where the concatenated code implementation 
only requires local operations of one or two modes (i.e.,~$N=1,2$),
computing the PTM+ model of each operation is thus possible.

Some intuition about the meaning of the individual matrix elements may be gained by fixing $e, e'$, and interpreting
the tensor $(S_{e, e'})_{\ell, \ell'}$ as a sub--normalized PTM for the logical channel associated to
the error section transition $e' \to e$.
The output population in sector $e$ is governed by the
coefficients $S_{e,e',I,\ell'}$.
When $\ptmp$ is acting on
a density matrix $\hat \rho = \sum \rho_{e'\ell'} \pauli_{e' \ell'}$, 
the output population in sector $e$ may depend both on the populations $\rho_{e'I}$ 
and the logical states in each sector expressed by $\rho_{e', \ell'\neq I}$.
The remaining coefficients $S_{e,e',\ell \neq I,\ell'}$ express the logical action applied alongside the $e'\rightarrow e$ error sector transition.

To understand the impact of the PTM+ approximation in the context of a larger simulation, consider a circuit $W = \channel_T \circ ... \circ \channel_1$.
Approximating each individual channel with PTM+ gives the approximated circuit
\begin{equation}
    W_\text{PTM+} = \PR \circ \channel_T \circ \PR \circ \channel_{T-1} \circ \PR \circ ... \circ \PR \circ \channel_1 \circ \PR,
\end{equation}
where we have used $\PR \circ \PR = \PR$. We assume that the initial and final quantum states are error--diagonal, which is trivially true if $\channel_1$ initializes each quantum mode and $\channel_T$ destructively measures each quantum mode. 
There are two extreme ways in which $W_\text{PTM+}$ can be a perfect approximation to the full circuit $W$.
Firstly, if each channel satisfies $\channel_t \circ \PR = \PR \circ \channel_t \circ \PR$, then $W_\text{PTM+} = W$.
This condition means that none of the channels generate coherence between error sectors, when acting on an error--diagonal state. 
In this case, the density matrix after each channel remains error--diagonal even in the full circuit $W$, and dropping coherences between error sectors has no impact.
However, this condition is not necessary:
If each channel instead satisfies $\PR \circ \channel_t = \PR \circ \channel_t \circ \PR$, and the final state is error--diagonal, then we likewise have $W_\text{PTM+} = W$.
This condition means that the error--diagonal part of the output state of $\channel_t$ is independent of the coherences between error sectors in the input state.
In this case, mid--circuit density matrices may feature coherences between error sectors, but these coherences never ``spill over'' into the error diagonal part of the state. In other words, in this case there is no interference between error sectors.
More realistically, in order for PTM+ to be a good approximation for a given circuit, a condition between the two extremes needs to be achieved:
The error--diagonal part of the output of each channel $\channel_t$ should not depend strongly on those
coherences between error sectors which are actually generated by the preceding channels $\channel_{<t}$.

\subsection{Bosonic Pauli+ channels}
While the action of PTM+ channels on the error Hilbert spaces of each involved mode is highly restricted, the action on the logical Hilbert spaces is still 
general. A simulation where error channels are represented with PTM+ would then still require exponential resources in the number of modes. 
To arrive at BP+ channels, we thus make a second approximation by applying Pauli twirling \cite{geller_efficient_2013}. This approximates the 
logical action of an error channel as a Pauli channel, which can be simulated with polynomial resources using stabilizer tableau methods \cite{gottesman_theory_1998, aaronson_improved_2004}.

The resulting BP+ channel is fully parameterized by a transition matrix $p({e|e'})$, with associated Pauli error rates $p({\ell | e, e'})$ to be applied alongside each error sector transition $e' \rightarrow e$.
The action on a general density matrix $\hat \rho$ can be written as
\begin{equation}\label{eq:bpplus_action_on_density_matrix}
 \bpp(\hat \rho) = \sum_{e, e'} p({e | e'})  \sum_\ell p({\ell | e , e'}) (\op{e}{e'} \otimes \pauli_\ell) \cdot \hat \rho \cdot (\op{e'}{e} \otimes \pauli_\ell).
\end{equation}
The transition matrix satisfies $\sum_e p({e|e'}) = 1$, and the Pauli error rates satisfy $\sum_\ell p(\ell | e,e') = 1$. 
The computation of the BP+ parameters, as a function of the PTM+ parameters, is outlined in appendix \ref{app:ptmp_to_bpp}. 
When a BP+ channel is written in PTM+ form, the PTM+ coefficients are ``logically diagonal'', i.e.~$S_{e,e',\ell,\ell'} \sim \delta_{\ell, \ell'}$.

\begin{figure*}[t]
    \centering
    \includegraphics[width=.7\textwidth]{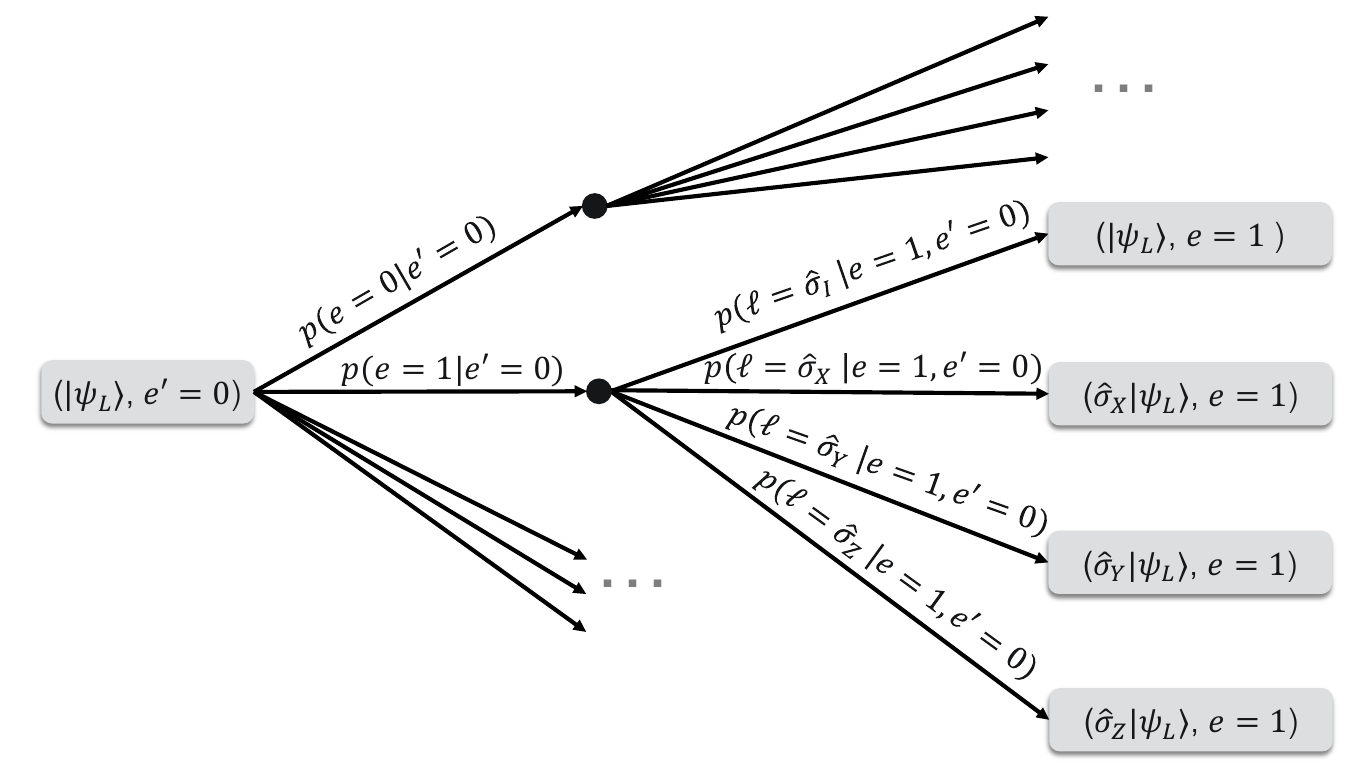}
    \caption{ {Illustration of the sampling--based application of a BP+ channel.
    A BP+ channel is described by an error sector transition matrix $p(e|e')$ and Pauli error rates $p(\ell | e, e')$. 
    Its action on an input state $\ket{\psi} = \ket{e'=0} \otimes \ket{\psi_L}$, with input error sector $e'=0$ and logical state $\ket{\psi_L}$, can be sampled from as follows: First, sample the output error sector $e$ according to $p(e|e')$. Then, sample which Pauli matrix to apply to the logical state, according to $p(\ell|e, e')$. Illustrated here is a channel of a single bosonic qubit, where there are four possible Pauli matrices.}}
    \label{fig:bpplus_mc_application}
\end{figure*}

A BP+ channel, like an ordinary Pauli channel, can be applied to pure states in a Monte--Carlo fashion. 
To see it, let $\ket\psi = \ket{e'}\otimes \ket{\psi_L}$ be a pure error--diagonal state, which must necessarily be 
a basis state on the error Hilbert space. 
The action of a BP+ channel on this state can be written as a nested expectation value, which can straightforwardly be sampled from:
\begin{equation}\label{eq:bpplus_mc_application}
    \bpp\left(\op\psi\right) = \mathbb E_{e \sim p(e|e')} \qty{ \mathbb{E}_{\ell \sim p(\ell|e, e')} \left(
    \op {\phi} \right)}, \text{where } \ket{\phi} = \ket{e} \otimes (\pauli_\ell \cdot \ket{\psi_L}).
\end{equation}
Here, $\mathbb{E}_{x \sim p(x)}$ denotes the expectation value where $x$ is sampled from $p(x)$. $\ket{\phi}$
is again a pure error--diagonal state, such that several BP+ channels can be sampled in sequence. 
 {Sampling from eq.~\eqref{eq:bpplus_mc_application} is illustrated in fig.~\ref{fig:bpplus_mc_application}.}

In order to use BP+ models for concatenated code simulations, we need to model non--Pauli 
Clifford gates, like the CNOT gate, which cannot directly be expressed by BP+.
Following standard practice, we model noisy implementations of a non--Pauli Clifford gate as an ideal Clifford gate, followed by a noise channel, and set $\channel = \channel_\text{noise} \circ \channel_\text{ideal}$.
Here, $\channel_\text{ideal} =  \mathcal I_E \otimes \channel_L$ is the ideal gate--level action of the desired Clifford gate on the logical Hilbert space, and acts 
trivially on the error Hilbert spaces. The ideal action is followed by $\channel_\text{noise} = \channel \circ (\channel_\text{ideal})^{-1}$ which we approximate as a BP+ channel.

We also want to devise BP+ models for operations which return a classical outcome.
An important example is the $\sBs$ stabilization protocol of the GKP code presented in section \ref{sec:sbs_basis}, which returns an inner--code syndrome measurement outcome.
To model this, we use the notion of quantum instruments \cite{davies_operational_1970}:
The action of a channel $\channel$ with a binary classical outcome on a density matrix $\hat \rho$ is described by two linear maps $\mathcal C_{0, 1} (\hat \rho)$, 
such that $\mathcal C_{0} + \mathcal C_1$ 
is a valid quantum channel. The probability of measuring outcome $o$ is given by $p_o = \tr \mathcal C_o(\hat \rho)$, and the conditional 
post--measurement state by $\mathcal C_o(\hat\rho) / p_o$.
We separately approximate the channels $\channel_{o}$ as PTM+ channels and then as BP+ channels. The PTM+ quantum instrument is then parameterized by coefficients $S_{o, e, e', \ell, \ell}$, and 
the BP+ model is parameterized with a transition matrix $p({o, e| e'})$, and Pauli error rates for each transition $p(\ell | o, e, e')$.

The effects of Pauli twirling can be understood by comparing the properties of PTM+ and BP+ channels. 
Firstly, as the names suggest, the logical action associated with each error space transition is general for 
PTM+ channels, and is restricted to a Pauli channel for BP+. 
Approximating errors as Pauli channels is a standard approximation in the field 
of quantum error correction, and can be motivated by the fact that 
outer--code stabilizer measurements approximately project more general errors to Pauli errors \cite{beale_quantum_2018}.
More general errors can also be transformed into Pauli errors  {in a practical circuit implementation} by randomized compiling techniques \cite{wallman_noise_2016}.

A second more subtle difference between the PTM+ and BP+ models relates to the interplay between error and logical Hilbert spaces:
As we saw, in PTM+, output populations in each error sector may depend on the input logical state in each sector, and not just the input populations in each sector, 
as expressed by the coefficients $S_{e, e', \ell=I, \ell'\neq I}$.
In BP+ however, output populations may depend on input populations only, as expressed by the transition rates $p(e|e')$.
Similarly, in PTM+ the probabilities of classical outcomes $o$ may depend on the logical state in each sector while in BP+ they can depend on input populations only.
In order for BP+ to be a good approximation for a channel $\channel$, we thus need that output populations and outcome probabilities of $\channel$ 
are approximately independent of the input logical states, for relevant input states.
This condition is important not just for the quality of modelling with BP+, but also for performance: 
If output error sector populations and outcome probabilities did depend on input logical states, 
logical information would leak into the error Hilbert space and measurement outcomes.
This would cause logical decoherence.

\subsection{Simulation of Clifford circuits with BP+ noise channels}
\label{sec:circ_sim}

\begin{algorithm}
\fbox{\parbox{.985\textwidth}{
\parbox{.975\textwidth}{
\textbf{Algorithm to sample evolution in error space}
\begin{itemize}
    \item Input: Circuit $W$ with depth $T$, represented as a list of operations $\channel_t$. Each operation is either an ideal Clifford operation, or a BP+ error model with or without classical outcome.
    \item Initialize empty lists $O$ for the outcomes of BP+ models, and $W_s$ for the sampling circuit.
    \item Initialize $\vec e = \vec 0$, containing the initial error sector index of every mode of the circuit.
    \item For every operation $\channel_t$ in the circuit $C$:
    \begin{itemize}
        \item If $\channel_t$ is an ideal Clifford operation, append $\channel_t$ to $W_s$.
        \item Else, $\channel_t$ is a BP+ channel represented by $(Q_t, p_t(o, e|e'), p_t(\ell|o, e, e'))$.
        Here $Q_t = \{q_i\}$ contains the indices of the modes which $\channel_t$ acts on.
        For the purpose of this algorithm, for uniform notation, we consider BP+ channel without an outcome to always return a trivial fixed outcome $o=0$.
        \begin{itemize}
            \item Set the input error sector indices $\tilde e'$ of the present channel $\channel_t$
            by selecting entries at indices $Q_t$ of $\vec e$
            \item Sample the outcome $\tilde o$ and output error sector indices $\tilde e$ from $p_t(o, e|e' =\tilde e')$
            \item Select the Pauli error rates $p_\ell = p_t(\ell| o = \tilde o, e=\tilde e, e'=\tilde e')$.
            \item Append the Pauli error channel with rates $p_\ell$, acting on the modes $Q_t$, to $W_s$
            \item Update the entries at indices $Q_t$ of $\vec e$ to the sampled output error sector indices $\tilde e$ of the present channel
            \item Append $\tilde o$ to $O$
        \end{itemize}
    \end{itemize}
    \item Return the sampling circuit $W_s$, and BP+ outcomes $O$.
\end{itemize}}}}
\caption{Algorithm to simulate the evolution of error sector indices and the BP+ outcomes, for a circuit which contains ideal Clifford operations and BP+ channels. The algorithm yields a random instance of the BP+ outcomes for every BP+ channel, and the sampling circuit, which is an ordinary noisy Clifford circuit from which the 
evolution of the logical quantum state can be sampled in a second step.}
\label{algo:classical_part}
\end{algorithm}

In this section we outline how a circuit containing Clifford operations and noise modelled by BP+ channels may be efficiently simulated.

We recall that ordinary Clifford circuits with two--level qubits are a class of circuits that are classically simulatable~\cite{gottesman_heisenberg_1998, aaronson_improved_2004} with resources polynomial in the number 
of qubits $N$. 
Clifford circuits include all operations typically used for QEC with qubit codes.
The quantum states which can arise in Clifford circuits can be efficiently represented using a stabilizer tableau $Z$, which in its most basic version is a 
matrix of size $(2N + 1) \cdot N$ with entries in $\{0, 1\}$. Clifford operations can be efficiently applied to this state.
Noise is typically simulated in a probabilistic fashion: a noisy implementation of a Clifford gate is modelled as an ideal Clifford gate, 
followed by a Pauli noise channel. The Pauli noise channel can be applied by randomly sampling a Pauli operator from a given probability distribution, and
applying it to the stabilizer tableau. The simulation of the whole circuit is then repeated many times to examine the average behaviour.

A straightforward implementation of a Monte--Carlo simulation of a circuit containing BP+ noise channels would then represent a pure quantum state by the data $(\vec e, Z)$. 
Here, $\vec e$ is a multi--index containing the error sector index $e_n$ for every mode $n$, 
and $Z$ is a stabilizer tableau representing the state on the multimode logical Hilbert space. 
An ideal Clifford channel can be applied to this data by processing $Z$ and leaving $e$ untouched, 
and a BP+ channel can be applied probabilistically following eq.~\eqref{eq:bpplus_mc_application}.

In practice however, we evolve the ``classical part'' $\vec e$ through the whole circuit first, 
and only then evolve the ``quantum part'' $Z$ of the state.
This is possible because in BP+, the transition amplitudes $p(e | e')$ and the 
probabilities of any outcomes $o$ do not depend on the logical state described by $Z$.
The algorithm for propagating $e$ and sampling the BP+ outcomes $o$ is given in algorithm \ref{algo:classical_part}.
After sampling the evolution $e(t)$ and outcomes $o(t)$ at every step $t$ of the circuit, 
the appropriate Pauli error rates $p\left(\ell | o=o(t), e =  e(t), e' =  e(t-1)\right)$ are selected from each BP+ channel and plugged into the circuit.
The algorithm thus returns an ordinary Clifford circuit with Pauli noise, which we refer to as the ``sampling circuit''.
In a separate second step, the evolution of the logical quantum state is sampled using the sampling circuit.
This two--step procedure has the advantage that standard tools can be used for the second step. It also ensures that Clifford simulation algorithms with improved efficiency compared to basic stabilizer tableau evolution can be applied.
An example of the data arising from the first step is given in section \ref{sec:sampled_error_space_dynamics}.

\section{The sBs Basis}\label{sec:sbs_basis}
In this section, we offer reminders about the GKP code and the small--big--small (sBs) protocol, introduce a new basis which we call the sBs basis, and describe the sBs protocol using the sBs basis.

\subsection{Background: the GKP code and the sBs protocol}
The GKP code \cite{gottesman_encoding_2001} encodes a logical qubit into the Hilbert space of a quantum harmonic oscillator.
The code states of the single--mode finite energy square GKP code, which we consider here, can be written as
\begin{align}\label{eq:analytic_fe_states}
    \ket{0_\Delta} ={} \frac{1}{\mathcal N_0} e^{-\Delta^2 \hat n} \sum_{k=-\infty}^\infty \ket{\hat q = k l}, \qquad
    % \nonumber\\
    \ket{1_\Delta} ={} \frac{1}{\mathcal N_1} e^{-\Delta^2 \hat n} \sum_{k=-\infty}^\infty \ket{\hat q = (k+\tfrac12) l}.
\end{align}
Here $\Delta$ is the finite--energy parameter, $\hat n$ is the Fock number operator, $l = 2\sqrt{\pi}$ gives the distance between position--space peaks of the GKP states,
and $\mathcal N_{0, 1}$ are normalization factors ensuring that the code states have unit norm.
The code states can also be understood as the simultaneous eigenvectors, of eigenvalue $+1$, of the two finite--energy stabilizer operators
\begin{align}
    \hat S^q_\Delta ={} e^{-\Delta^2 \hat n} e^{i l \hat q} e^{\Delta^2 \hat n}, \qquad % \nonumber\\
    \hat S^p_\Delta ={} e^{-\Delta^2 \hat n} e^{-i l \hat p} e^{\Delta^2 \hat n},
\end{align}
where $\hat q$ and $\hat p$ are the position and momentum operators of the bosonic mode.

Physical errors, like photon losses of the quantum oscillator hosting the GKP state, can cause the state to leave the code space. The GKP encoding allows the implementation of an operation which restores states that have suffered physical errors to the code space, in a way that preserves the logical information.
The leading proposal to approximately achieve this in the superconducting setting is the sBs protocol, introduced in refs.~\cite{ de_neeve_error_2022, royer_stabilization_2020} and experimentally implemented in refs.~\cite{lachance-quirion_autonomous_2023, sivak_real-time_2023}.

\begin{figure}
\centering
\includegraphics{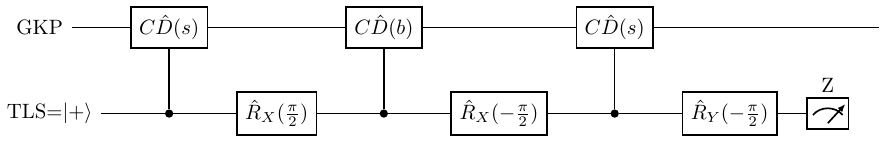}
\caption{Gate sequence for $\sBs$ stabilization. In order to implement $\sBs_q$, we set $s = \frac12 \sinh(\Delta^2) l$ and $b = -i \cosh(\Delta^2) l$. For $\sBs_p$, we set $s = \frac i 2 \sinh(\Delta^2) l$ and $b = \cosh(\Delta^2) l$.
Here ${C\hat D}$ are conditional displacement gates defined in eq.~\ref{eq:ideal_cd}, and $\hat R_{P}(\phi) = \exp(- \frac i2 \phi \hat P)$ are rotations around the $\hat P$ axis of the TLS Bloch sphere, where $\hat P \in \{\hat X, \hat Y, \hat Z\}$.}
\label{fig:sbs_circuit}
\end{figure}

The $\sBs$ protocol involves an auxiliary TLS in addition to the bosonic mode hosting the GKP state, and is depicted in fig.~\ref{fig:sbs_circuit}.
The only entangling gate required is the conditional displacement gate ${C\hat D}(\beta)$
between a TLS and 
a bosonic mode, which performs a displacement by $\pm \frac{\beta}{2}$ of the bosonic mode, where the sign of the 
displacement depends on the state 
of the TLS. In absence of errors, the unitary action of the gate reads
\begin{equation}\label{eq:ideal_cd}
    {C\hat D}(\beta) = \exp\left( \tfrac{1}{2\sqrt{2}} (\beta \hat a ^\dagger - \beta^* 
    \hat a) \otimes \hat \sigma_z \right),
\end{equation}
where $\hat a$ is the annihilation operator of the bosonic mode, and $\pauli_z$ pertains to the TLS.

The  {space of $+1$ eigenvectors} of $\hat S^q_\Delta$ and $\hat S^p_\Delta$ are stabilized by two separate protocols, 
which we term $\sBs_q$ and $\sBs_p$ respectively. After a physical error has acted on the state, 
several rounds of stabilization may be required to restore a state to the  {space of $+1$ eigenvectors}, and in practice, one alternates between stabilization of the two quadratures.
The $\sBs_q$ protocol additionally applies a logical $Z$ operation, and the $\sBs_p$ protocol applies a logical $X$ operation.
In the version of the protocols used here, finding the TLS in its $\ket{1}$ state at the end of the protocol 
approximately indicates that an oscillator error has been 
corrected, as observed in ref.~\cite{sivak_real-time_2023} and explored in more detail in section \ref{subsec:sbs_subspace_decomp}.

If no errors occur during the implementation of the protocol, the ideal action of $\sBs_q$ on a density matrix $\hat \rho$ of the GKP mode can be described by two Kraus operators 
$\hat K_0^q, \hat K_1^q$, which are obtained by evaluating the circuit of fig.~\ref{fig:sbs_circuit} without noise. The probability of measuring $o \in \{0, 1\}$ is given by $p_o = \tr(\hat K_o^q{}^\dagger \hat K_o^q{} \hat \rho)$, and the final 
state of the GKP mode conditioned on having measured $o$ is given by $\hat K_o^q \hat \rho \hat K_o^q{}^\dagger / p_o$, and similarly for $\sBs_p$.
We emphasize that the Kraus operators act on the bosonic mode only, and not on the auxiliary TLS. Analytic expressions for the Kraus operators are given in ref.~\cite{royer_stabilization_2020}.

\subsection{The sBs basis}\label{subsec:sbs_subspace_decomp}

\begin{algorithm}
\fbox{\parbox{.985\textwidth}{
\parbox{.975\textwidth}{
    \textbf{Algorithm to construct no--error states}
    \begin{itemize}
        \item Construct the projector onto the two eigenvectors with largest eigenvalue of the operator $\frac12 (\hat K_0^q{}^\dagger \hat K_0^q + \hat K_0^p{}^\dagger \hat K_0^p)$
        \item Project the states of eq.~\eqref{eq:analytic_fe_states} using that projector, and orthonormalize using L\"owdin's procedure.
        \item Return the resulting states as $\{\ket{(0,0),0}, \ket{(0,0),1} \}$.
    \end{itemize}
    \textbf{Algorithm to construct the sBs basis}
    \begin{itemize}
        \item Input: Maximum error rank $R$, operators $\hat K_1^q{}^\dagger$ and $\hat K_1^p{}^\dagger$, and orthonormal no--error states $\{\ket{(0,0), 0}, \ket{(0,0), 1} \}$, in the bosonic Hilbert space of even dimension $d$
        \item Initialize $B_{\sBs} = \{\ket{(0,0), 0}, \ket{(0,0), 1} \}$
        \item For each error rank $r$ between $1$ and $R$ (inclusive), do:
        \begin{itemize}
            \item Initialize a list of candidate vectors for the error rank $r$: $\tilde B_{\sBs}^{r} = \{\}$
            \item For each error sector of rank $r$ (i.e., for each tuple $(e_q \ge 0, e_p \ge 0)$ with $e_q + e_p = r$), construct the candidate vectors for that error sector as follows:
            \begin{itemize}
                \item If $e_p>0$, obtain the $p$--route candidates as $\ket{(e_q, e_p), \mu}_p = \hat K_1^p{}^\dagger \ket{(e_q, e_p-1), 1-\mu}$. The states on the r.h.s.~are taken from $B_{\sBs}$, and we have compensated for the logical $\pauli_x$ action of $\sBs_p$ by swapping $\mu = 0\leftrightarrow 1$. Else (i.e.~if $e_p=0$), set the $p$--route candidates to $0$.
                \item If $e_q>0$, obtain the $q$--route candidates as $\ket{(e_q, e_p), \mu}_q = (-1)^\mu \hat K_1^q{}^\dagger \ket{ (e_q-1, e_p), \mu}$, where $(-1)^\mu$ compensates the logical $\pauli_z$ action of $\sBs_q$. Else (i.e.~if $e_q=0$) set the $q$--route candidates to $0$.
                \item Set the candidate vectors $\ket{(e_q, e_p), \mu}$ as the phase--corrected average (defined below) of $\ket{(e_q, e_p), \mu}_{q}$ and $\ket{(e_q, e_p), \mu}_{p}$.
                \item Append $\ket{(e_q, e_p), 0}$ and $\ket{(e_q, e_p), 1}$ to $\tilde B_{\sBs}^{r}$.
            \end{itemize}
            \item Project each vector in $\tilde B_{\sBs}^{r}$ onto the complement of the span of $B_{\sBs}$, and normalize vectors individually.
            \item Orthonormalize the members of $\tilde B_{\sBs}^{r}$ amongst each other using L\"owdin's procedure
            \item Append the members of $\tilde B_{\sBs}^{r}$ to $B_{\sBs}$.
        \end{itemize}
        \item Generate $d - |B_{\sBs}|$ random complex Gaussian vectors, project them onto the complement of the span of $B_{\sBs}$, orthonormalize them amongst each other, and append them to $B_{\sBs}$.
        \item Return $B_{\sBs}$.
    \end{itemize}
    \textbf{Phase--corrected average}
    \begin{itemize}
        \item Input: Vectors $\ket{e, 0}_q$, $\ket{e, 0}_p$, $\ket{e, 1}_q$, $\ket{e, 1}_p$. 
        \item Find a phase $\phi \in \mathbb{R}$ such that $\sum_{\mu\in\{0, 1\}} \bra{e, \mu}_q e^{i\phi} \ket{e, \mu}_p$ is real and non--negative.
        \item Return $\ket{e, \mu} = \frac12 \left(\ket{e, \mu}_q + e^{i\phi} \ket{e, \mu}_p\right)$.
    \end{itemize}
}}}
    \caption{Algorithms to construct the sBs basis.}
    \label{algo:sBs_basis_algo}
\end{algorithm}

This section introduces the basis which we use to decompose bosonic Hilbert space into subsystems, and to extract PTM+ and BP+ models from numerical propagators. We refer to the basis constructed here as the ``sBs basis''.
As we saw, the quality of the approximations made by BP+ depends on the choice of basis. 
In particular, BP+ neglects decoherences between the error sectors defined by the basis,
and we thus want to build a basis such that these coherences do not play an important role in the dynamics  {during error correction}.

We start from the observation that an $\sBs_{q/p}$ measurement outcome of $o=1$ indicates that probably,
an error has been corrected by applying $\hat K_1^{q/p}$.
One may thus think of $\hat K_1^q$ as destroying one type of error, 
and $\hat K_1^p$ destroying another type of error. 
By analogy to the ladder operators of a harmonic oscillator, 
this suggests that $\hat K_1^q{}^\dagger$ and $\hat K_1^p{}^\dagger$ can be used to 
artificially create errors. 
Using this intuition, a basis can be built by starting from a two--dimensional basis of finite--energy logical ``no--error'' GKP states, and interpreting them as the first two elements $\ket{e=0, l=0}$ and $\ket{e=0, l=1}$. By acting with operators of the form $(\hat K_1^q{}^\dagger)^{e_q} \cdot (\hat K_1^p{}^\dagger)^{e_p}$ on the first 
two elements, we then obtain the basis elements $\ket{e=(e_q, e_p), l=0}$ and $\ket{e=(e_q, e_p), l=1}$.
An error sector is thus indexed by two integers $e = (e_q, e_p)$, indicating the number of 
$\hat K_1^q{}^\dagger$-like and $\hat K_1^p{}^\dagger$-like errors. We refer to $r=e_q + e_p$ as the 
rank of an error sector.

To arrive at a complete algorithm for building the $\sBs$ basis, a few more choices need to be made.
Firstly, a precise choice of no--error states needs to be made. 
Empirically, we found that using the analytic finite--energy states of eq.~\eqref{eq:analytic_fe_states} does not make BP+ a good approximation.
Instead we use the 2D sector of states which maximizes the probability of measuring $o=0$ when performing gate--level $\sBs$, averaged over applying $\sBs_q$ or $\sBs_p$.
This choice takes into account that because $\sBs$ only approximately stabilizes finite--energy GKP states, there is a subtle difference between the analytic states eq.~\eqref{eq:analytic_fe_states} and the states most likely to give a trivial $\sBs$ outcome.
 {The squared fidelity between the states of eq.~\eqref{eq:analytic_fe_states} and the no--error states of the $\sBs$ basis is $0.995$ -- $0.997$ depending on logical state at $\Delta=0.36$, and increases with decreasing $\Delta$.}

Secondly, one needs to ensure that the constructed basis is orthonormal. We achieve this here by 
extending the basis rank by rank. Candidates for the basis elements in rank $r+1$ are constructed from the basis elements 
in rank $r$ by acting with $\hat K_1^q{}^\dagger$ or $\hat K_1^p{}^\dagger$, and are orthonormalized to the previous basis elements and amongst each other before being 
added to the basis. We use L\"owdin's symmetric orthonormalization procedure \cite{lowdin_nonorthogonality_2004}
to minimize the changes to the candidate vectors caused by the orthonormalization.
For some error sectors, there can be two routes that may be taken to construct the error sector: For example, the error sector $e=(1,1)$ may be obtained by acting with $\hat K_1^q{}^\dagger$ on $e=(0,1)$, or with $\hat K_1^p{}^\dagger$ on $e=(1, 0)$. Here we take a phase--corrected average of the two routes, fixing an unimportant global phase ambiguity.

Thirdly, the operators $\hat K_1^q{}^\dagger$ and $\hat K_1^p{}^\dagger$ also have the effect of applying a logical Pauli operator, which needs to be undone when using these operators to raise the error sector index, in 
order to ensure that the logical content of each basis vector $\ket{e, \mu}$ is independent of $e$.
Lastly, the algorithm presented here may fail when the size of the basis approaches the dimension of the bosonic truncated Hilbert space,
since close to the Fock cutoff, the candidates for the next rank may fail to be linearily independent. 
As a simple way to circumvent this, we choose a maximum rank
which leads to a basis which is not complete (has fewer vectors than the Hilbert space dimension), and ``fill up'' the basis 
with suitably orthonormalized random vectors.
While the PTM+ and BP+ approximations are not expected to be accurate 
when the random basis elements are involved,
most of the population resides in the lower error sectors as shown in appendix \ref{app:additional_sims}, 
such that predictions are not strongly affected by dynamics close to the Fock cutoff.
A complete algorithm is presented in algorithm \ref{algo:sBs_basis_algo}.

\begin{figure*}[t]
    \centering
    \includegraphics[width=\textwidth]{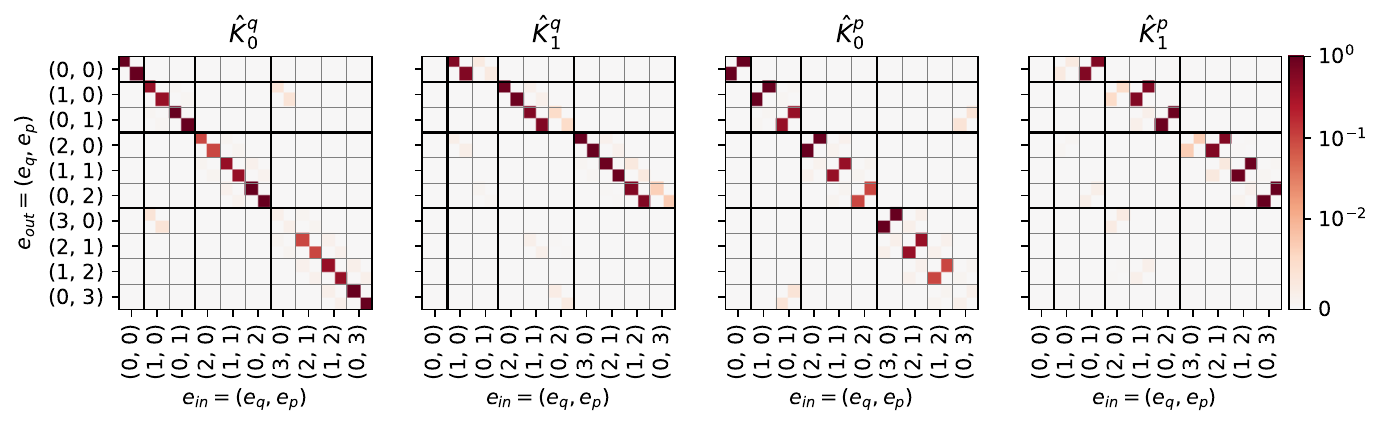}
    \caption{The squared absolute values of matrix elements of the four Kraus operators $\hat K_{0,1}^{q, p}$, in the $\sBs$ basis, truncated to error sectors of the first four ranks.
    The color scale has been chosen to emphasize small matrix elements, and is linear between $0$ and $10^{-2}$ and logarithmic between $10^{-2}$ and $1$.
    }
    \label{fig:sbs_kraus}
\end{figure*}

\subsection{sBs stabilization in the sBs basis}
Figure \ref{fig:sbs_kraus} shows the matrix elements of the four 
Kraus operators $\hat K_{0, 1}^{q, p}$, in the $\sBs$ basis.
We see that indeed, $\hat K_1^q$ has the effect of lowering the number of $\hat K_1^q{}^\dagger$--like errors $e_q$ by one, 
up to small subdominant matrix elements, and similarly $\hat K_1^p$ lowers $e_p$ by one.
This justifies, a posteriori, our intuition of two independent types of errors, for which $\hat K^q_1$ and $\hat K^p_1$ act as ladder operators.
Moreover, $\hat K_0^{q}$ and $\hat K_0^p$ are mainly supported on the block-diagonal, 
and cause little mixing between error sectors, with exceptions for some error sectors.

 {From fig.~\ref{fig:sbs_kraus}, it is also apparent that the probability of measuring $1$ (i.e., the norm of $\hat K_1^{q/p} |\psi\rangle$) differs for different input error sectors. The outcome history of repeated $\sBs$ rounds thus allows an inference about error sector populations at a given point in time, and repeated $\sBs$ can be seen as an approximate weak measurement of the error sector index.}

It is remarkable how sparse the $\sBs$ Kraus operators are,
when they are written in the $\sBs$ basis.  {This sparsity promises to make the PTM+ model a good approximation, when simulating the $\sBs$ protocol in the $\sBs$ basis.
Concretely, the Kraus operators mostly have one dominant element on each column, such that they do not generate coherences between error sectors, when acting on an error--diagonal state. In the notation of section \ref{sec:ptmp_channels}, we have $\sBs \circ \mathcal P_{\mathcal R} \approx \mathcal P_{\mathcal R} \circ \sBs \circ \mathcal P_{\mathcal R}$.
Additionally, the Kraus operators mostly have one dominant element on each row, such that there is little opportunity for constructive or destructive interference if the input state contains a superposition of error sectors. In other words, we have 
$\mathcal P_{\mathcal R} \circ \sBs \approx \mathcal P_{\mathcal R} \circ \sBs \circ \mathcal P_{\mathcal R}$. Altogether, the conditions of section \ref{sec:ptmp_channels} for the PTM+ model to be a good approximation are approximately satisfied when considering a circuit consisting of repeated rounds of $\sBs$.
The sparsity structure also distinguishes the $\sBs$ basis from the error subspace decomposition presented in ref.~\cite{sivak_real-time_2023} (appendix S4).
}

\section{BP+ models for sBs stabilization and CNOT gates}\label{sec:gates}
We now turn to applying the BP+ simulation method to a concatenated GKP code implementation with a specific noise model. This section
defines the noise model underlying the time evolution simulations, 
and extracts PTM+ and BP+ models for the required operations from the time evolution simulations.

\subsection{Noise model}\label{sec:noise_model}

\begin{table}
    \centering
    \begin{tabular}{c|c|c|c|c||c|c|c}
        $T_1^s$ & $T_\phi^s$ & $T_1^q$      & $T_\phi^q$ & $T_{ECD}$ & $d = |\hilbert_B|$ & $\Delta$ & Maximum error rank $R$\\
        \hline
        $1$ ms  & $100$ ms   & $100$ $\mu$s & $1$ ms & $500$ ns  & 196                & $0.36$   & $12$                  \\
    \end{tabular}
    \caption{Parameters of the noise model, the finite--energy GKP code, and the Fock cutoff used for numerical simulations.}
    \label{tab:noise_rates}
\end{table}

The only required entangling gate for the hybrid concatenated code simulated here is the conditional displacement gate eq.~\eqref{eq:ideal_cd} between a bosonic mode and a TLS.
This gate can be used to implement the $\sBs$ protocol shown in fig.~\ref{fig:sbs_circuit}. It is also the main constituent of the CNOT gates between the two--level outer--code syndrome qubits and the GKP data qubits as shown in section \ref{sec:cnot}. In this subsection, we introduce a physical noise model
for the CD gate.
Single--mode gates of the syndrome and auxiliary TLS are here modelled as instantaneous and perfect, which is justified since they are typically much faster and reach higher fidelity than the entangling operations.
Preparation and measurement of all modes are here modelled as noiseless for simplicity, but could also be included in BP+ models.

The CD gate can be implemented in superconducting hardware using the echoed conditional displacement (ECD) protocol \cite{eickbusch_fast_2022}. 
Alternative proposals to implement conditional displacements between a TLS and an oscillator exist 
\cite{touzard_gated_2019, puri_stabilized_2019}, and may also be amenable to modelling using BP+.
Here we model the effect of the main physical decoherence mechanisms on the implementation of the ECD protocol, 
while discarding any unitary implementation errors. Our model is
\begin{equation}\label{eq:ecd_model}
    \widetilde{CD}(\beta) = X \circ \widetilde{CD}_{\frac12} (-\beta/2) \circ X \circ \widetilde{CD}_{\frac12}(\beta/2).
\end{equation}
Here $X$ is the Pauli--X gate on the auxiliary TLS, and is modelled as a perfect gate. $\widetilde{CD}_{\frac12}(\beta)(\hat\rho)$ is defined as the solution at time $T_{ECD}/2$ 
of the Lindblad master equation
$
    \dot \rho = -i [\hat H_{CD}, \hat \rho] + \sum_i \mathcal D (\hat c_i) (\hat \rho)
$.
Here, 
\begin{equation} 
 \hat H_{CD}(\beta, T_{ECD}) = \frac{1}{T_{ECD}/2} \cdot \frac{1}{2\sqrt{2}} (i \beta \hat a^\dagger - i \beta^* \hat a) \otimes \hat \sigma_z
\end{equation}
is the conditional displacement Hamiltonian implementing the desired action (cf.~eq.~\eqref{eq:ideal_cd}).
The terms 
$\mathcal D(\hat c_i)(\hat \rho) = \hat c_i \hat \rho \hat c_i^\dagger - \frac12 \hat c_i^\dagger \hat c_i \hat \rho - \frac12 \hat \rho \hat c_i^\dagger \hat c_i$
express the impact of decoherence, and we include the collapse operators
\begin{align}
    \hat c_1 ={}& (T_1^q)^{-\frac12} \hat \sigma^-  & 
    \hat c_2 ={}& \sqrt{2} (T_\phi^1)^{-\frac12} (1 - \hat \sigma^z)/2 \nonumber \\
    \hat c_3 ={}& (T_1^s)^{-\frac12} \hat a &
    \hat c_4 ={}& \sqrt{2} (T_\phi^s)^{-\frac12} \hat a^\dagger \hat a.
\end{align}
$\hat c_1$ corresponds to decay of the TLS during the operation of the CD gate, with a lifetime $T_1^q$ of the excited state $\ket {1_q}$. 
Such a decay can cause large erroneous displacements of the 
bosonic mode, and since the GKP code does not protect well against large displacements, it can introduce a logical error. 
This effect is an important limitation of existing GKP implementations.
$\hat c_3$ corresponds to a decay of the 
bosonic mode, with a lifetime $T_1^s$ of the single--photon Fock state $\ket{1_s}$. 
Such a decay event is approximately correctable by the GKP code. 
$\hat c_2$ and $\hat c_4$ correspond to pure dephasing of the TLS and bosonic mode, respectively, 
with the phase coherence time given as $T_{2}^{q, s} = ( 1/T_\phi^{q, s} + 1/2 T_1^{q,s })^{-1}$.
 {For our simulations, we choose noise rates and an energy parameter $\Delta$ which are representative of existing hardware implementations. }
The noise rates, energy parameter $\Delta$ of the GKP code, 
as well as the simulation parameters which are used for our simulations, are given in table \ref{tab:noise_rates}.

While not all error channels which are expected in an experimental implementation have been included in our time evolution models,
the model used here presents an increase in realism over idealized models such as random 
Gaussian displacement noise, 
or Kraus operator noise applied in between rounds of stabilization.
We stress that BP+ is compatible with more fine--grained  {Markovian} noise models, 
such as pulse--level simulations, inclusion of additional dissipators,
and noisy models for preparation, measurement, idling, and single--mode operations.

\subsection{sBs stabilization}

\begin{figure}
    \centering
    \includegraphics[width=0.49\linewidth]{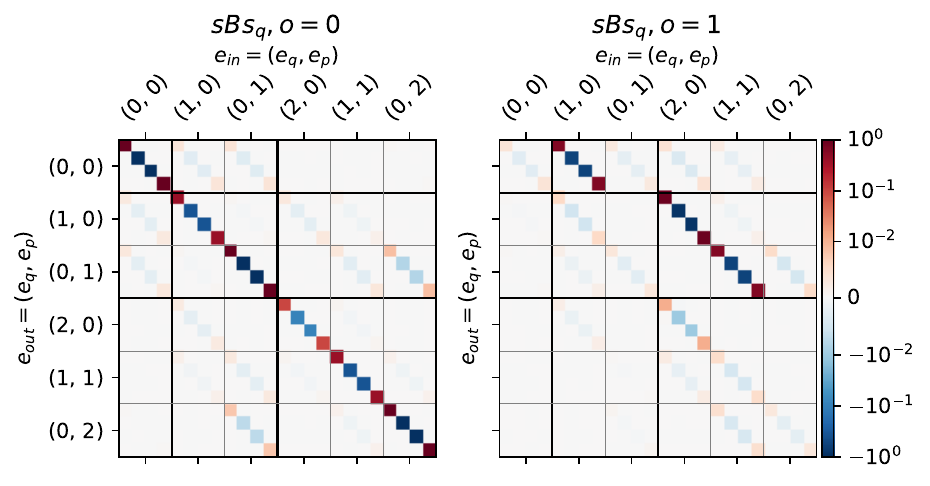}
    \includegraphics[width=0.49\linewidth]{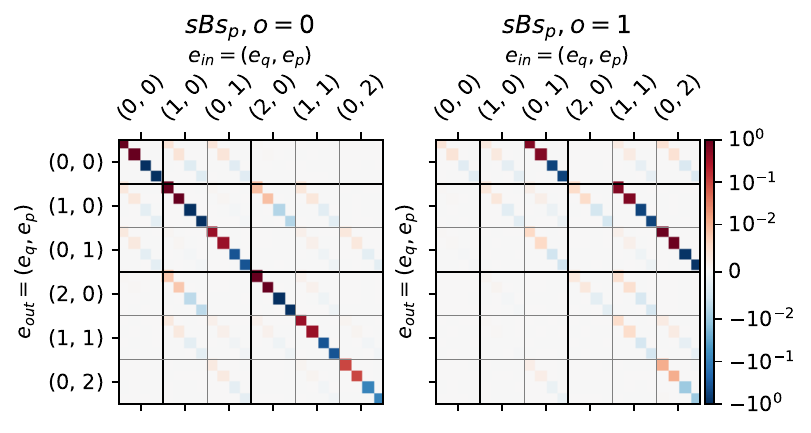}
    \caption{PTM+ data for sBs stabilization of a GKP state, extracted from a time evolution simulation with the noise rates listed in table \ref{tab:noise_rates}.
Shown are the coefficients $S_{o, e, e', \ell, \ell'}$ of the PTM+ models for $\sBs_q$ and $\sBs_p$. The sub--tensors corresponding to the measurement outcomes
$o=0$ and $o=1$ are shown separately, and are truncated to the first few error sectors. 
Rows are indexed by the ``output indices'' $e = (e_q, e_p)$ indexing the output error sector and $\ell$ indexing the output Pauli operator, while columns are indexed by the ``input indices'' $e'$ and $\ell'$.
There is one four--by--four block for each combination of $e$ and $e'$. Each four--by--four block can be interpreted as a single--qubit Pauli transfer matrix, 
with $\ell$ and $\ell'$ indexing entries of the four--by--four block.
The color scale is chosen to emphasize small matrix elements, and is linear between $-10^{-2}$ and $10^{-2}$ and logarithmic outside this interval.}
    \label{fig:stab_ptms}
\end{figure}

$\sBs$ stabilization is carried out using the circuit of fig.~\ref{fig:sbs_circuit}, with the noise model eq.~\eqref{eq:ecd_model} for the CD gates, and is simulated using a numerical Lindblad master equation solver \cite{johansson_qutip_2013}.
We extract the PTM+ models of stabilization in either quadrature using eq.~\eqref{eq:coeff_ptm_trace}. Heatmaps of the resulting matrices, for the first few error sectors, are shown in fig.~\ref{fig:stab_ptms}.
We can read off that for low--lying error sectors, the logical action is mostly as expected.
If the input state is in a higher error sector, then the $\sBs$ outcome $o=1$ becomes more likely, which is mostly accompanied by a decrease of the error space index as expected.
With small probability, the error sector index may change in other ways. This is partially due to decoherence, and partially because in the $\sBs$ basis, the error sectors are not perfectly un--mixed, as visible also from the presence of subdominant matrix elements in the Kraus operators shown in fig.~\ref{fig:sbs_kraus}. 
For low--lying error sectors, the logical PTM associated with each error sector transition is also mostly diagonal. This indicates that the logical action is approximately a Pauli channel, and that $\sBs$ outcome probabilities and error sector transition rates do not depend strongly on the logical state. For these reasons, Pauli twirling is a good approximation in low--lying error sectors.

\subsection{CNOT}\label{sec:cnot}

\begin{figure}
\centering
\begin{tikzpicture}
    \draw (0, 0) node[inner sep=0] {
\includegraphics[width=0.48\textwidth]{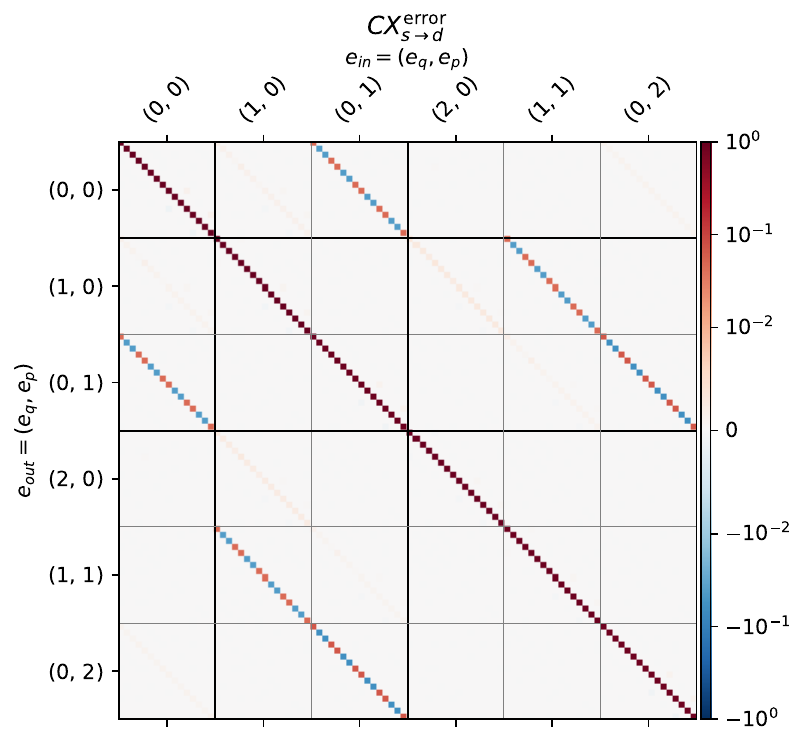}
\includegraphics[width=0.48\textwidth]{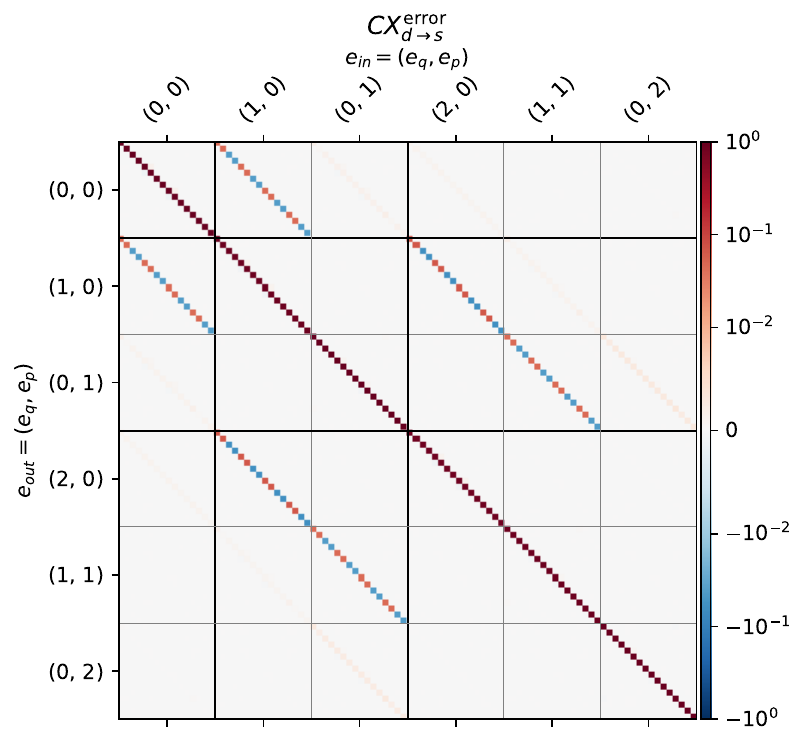}
};
\draw (-7, 3) node {\small\textsf{a)}};
\draw (0.5, 3) node {\small\textsf{b)}};
\end{tikzpicture}
\caption{PTM+ data for the error channel that accompanies the CNOT implementations of eq.~\eqref{eq:cx_implementations}.
a) Error channel that accompanies $CX_{s\rightarrow d}$, which has the syndrome qubit as the control qubit. b) Error channel that accompanies $CX_{d\rightarrow s}$, for which the GKP data qubit is the control qubit.
In either case, the CNOT implementations are modelled as $CX = CX^\text{error} \circ CX^\text{ideal}$, where $CX^\text{ideal}$ is the ideal action of the gate on the logical Hilbert space, and acts trivially on the error Hilbert space.
Each 16x16 block can be interpreted as two--qubit Pauli transfer matrix on the logical Hilbert space, and is thus indexed by the 16 Pauli strings $(II, IX, ..., ZZ)$, where the GKP mode is the first mode, and the syndrome TLS is the second mode.
The color scale has been chosen to emphasize small matrix elements, and is linear between $-10^{-2}$ and $10^{-2}$ and logarithmic outside this interval.}
\label{fig:cnot_ptm}
\end{figure}

To implement outer--code parity checks, CNOT gates between the GKP data qubits and two--level syndrome qubits are required. The syndrome qubit should act as the control of the 
CNOT for $X$ parity checks, and as its target for $Z$ parity checks.

Here, for the purpose of demonstrating the BP+ method, these gates are implemented using a single CD gate between the transmon and GKP qubit:
\begin{align}\label{eq:cx_implementations}
 CX_{s\rightarrow d} = \widetilde{CD} (\sqrt{\pi}), \qquad  CX_{d\rightarrow s} = H_s \circ \widetilde{CD} (- i \sqrt{\pi}) \circ H_s.
\end{align}
Here $CX_{s\rightarrow d}$ is the CNOT with the syndrome qubit as 
the control. $CX_{d\rightarrow s}$ has the data qubit as its 
control. 
It is implemented by sandwiching the gate 
$CZ_{s\rightarrow d} = \widetilde{CD}(-i\sqrt{\pi})$, 
which approximately implements the CZ gate with the TLS as the control, between two syndrome qubit Hadamard gates $H_s$.
The $CZ_{s\rightarrow d}$ gate can be obtained from the $CX_{s\rightarrow d}$ gate by sandwiching it between Hadamard gates on the GKP mode, 
which is equivalent to rotating the direction of the CD gate by $\pi/2$.

We decompose the action of the CNOT implementations as $CX_{s\rightarrow d} = CX_{s\rightarrow d}^\text{error} \circ CX_{s\rightarrow d}^\text{ideal}$, and similarly for $CX_{d\rightarrow s}$. In fig.~\ref{fig:cnot_ptm}, the PTM+ coefficients for $CX_{s\rightarrow d}^\text{error}$ and $CX_{d\rightarrow s}^\text{error}$ are shown.
We see from the diagonal blocks that the CNOT works well for low--lying error sectors, in the event that the error index remains constant.
However, for the gate $CX_{s\rightarrow d}$, there is a small probability the error space index $e_p$ changes up or down, and this is correlated with an additional $Z$ operation acting on the syndrome qubit.
Similarly, for $CX_{d\rightarrow s}$, the error space index $e_q$ may change up or down, which is correlated with an additional $X$ operation acting on the syndrome qubit.
These effects arise because a single CD gate only approximates a CNOT gate for a finite--energy GKP code, with the approximation becoming better in the limit $\Delta \rightarrow 0$ as the energy of the GKP state increases towards infinity.
For low--lying error sectors, the logical PTM associated with each error sector transition is also mostly diagonal, 
indicating that Pauli twirling is a good approximation.

In addition, the CNOT gates have the effect of displacing the peaks of the finite energy GKP 
state by half a lattice spacing, such that after the CNOT gate, the GKP state is no longer in the original code space.
As outlined in refs.~\cite{royer_encoding_2022, conrad_gottesman-kitaev-preskill_2022} in the context of multi--mode GKP codes, this can be compensated in software, 
by allowing the finite--energy stabilizers of the GKP code to take the value $-1$, and updating subsequent stabilization rounds and gates accordingly.
We detail this in appendix \ref{app:gauge}.

\section{Gaining Confidence in Model Accuracy}\label{sec:gaining_confidence}

While there are motivations and supporting theoretical arguments for the approximations we made, 
these arguments do not constitute a proof that BP+ is a good approximation in the context of simulating concatenated  {GKP} codes.  {In particular, the $\sBs$ basis in which the BP+ channels are expressed is built with the noiseless $\sBs$ protocol in mind, while realistic circuits also involve decoherence and CNOT--gates.} In this section, we thus gather empirical evidence for the quality of the BP+ approximations.

\begin{figure}
\centering
	\includegraphics{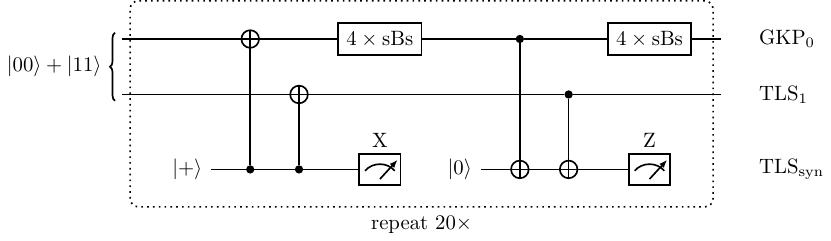}
    \caption{Circuit implementation of the prototypical two--qubit code used to compare predictions of time evolution simulations, and the PTM+ and BP+ models. One of the data qubits is simulated as a GKP qubit, and four rounds of stabilization are carried out after each CNOT in the sequence $(\sBs_q, \sBs_p, \sBs_q, \sBs_p)$.
    The other data qubit $\text{TLS}_1$ and the syndrome qubit $\text{TLS}_{\text{syn}}$ used to read out the qubit code stabilizers are modelled as two--level systems.
    The whole circuit is repeated 20 times.}
    \label{fig:XX_ZZ_code}
\end{figure}

\begin{figure}[t]
    \centering
\begin{tikzpicture}
    \draw (0, 0) node[inner sep=0] {\includegraphics[width=\linewidth]{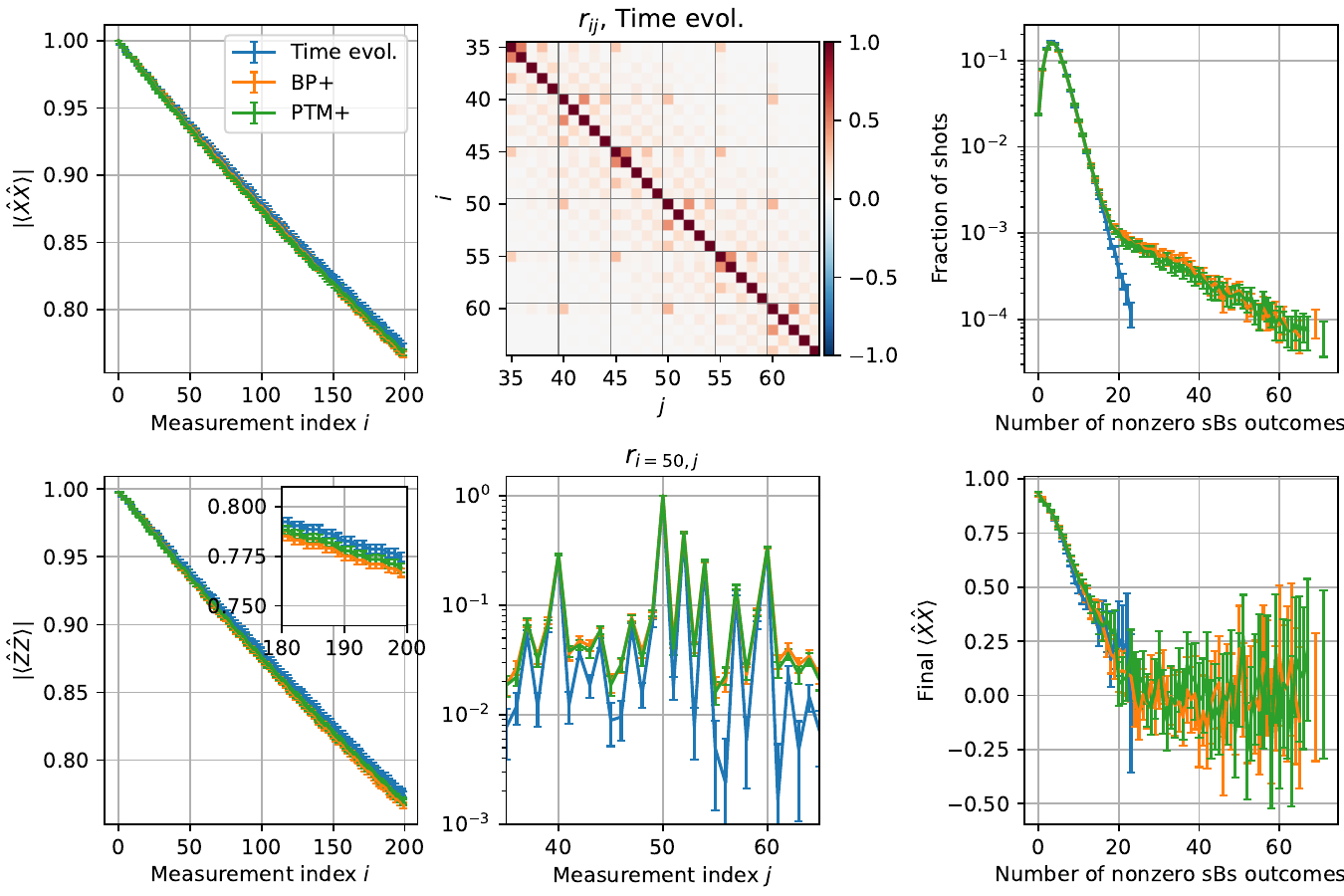}};
    \draw (-7, 5) node {\small \textsf{a)}};
    \draw (-7, 0) node {\small \textsf{b)}};
    \draw (-2.5, 5) node {\small \textsf{c)}};
    \draw (-2.5, 0) node {\small \textsf{d)}};
    \draw (2.9, 5) node {\small \textsf{e)}};
    \draw (2.9, 0) node {\small \textsf{f)}};
 \end{tikzpicture}
    \caption{
    Simulations of the two--qubit code described in section \ref{sec:gaining_confidence}. Measurement indices $10\cdot k, k \in \mathbb N_0,$ correspond to outcomes of the 
    measurement of the $\hat X \hat X$ outer--code stabilizer. Round indices $10 \cdot k + 5$ correspond to measurement of the $\hat Z \hat Z$ outer--code stabilizer. 
    Each outer--code measurement is followed by the sequence $(\sBs_q, \sBs_p, \sBs_q, \sBs_p)$ acting on the GKP qubit, which accounts for the remaining measurement indices.
    a), b): Expectation values of the outer--code stabilizers after each measurement, for each of the three models.
    c): Heatmap of the Pearson correlation coefficients between measurement outcomes indexed $i$ and $j$, for the time evolution model.
    d): Slice through c) at $i=50$, for all three models.
    e): Fraction of Monte--Carlo shots which have exactly $k$ non--zero sBs outcomes over the course of the circuit, for varying $k$.
    f): Expectation value $\langle \hat X \hat X \rangle$ after the final measurement, post--selected on having had exactly $k$ non--zero sBs outcomes over the whole circuit.
    Confidence intervals for all figures are 95\% confidence intervals obtained using the bootstrap method \cite{tibshirani_introduction_1994}, and in panels e) and f), points with fewer than ten corresponding samples have been discarded.
    }
    \label{fig:XX_ZZ_results}
\end{figure}
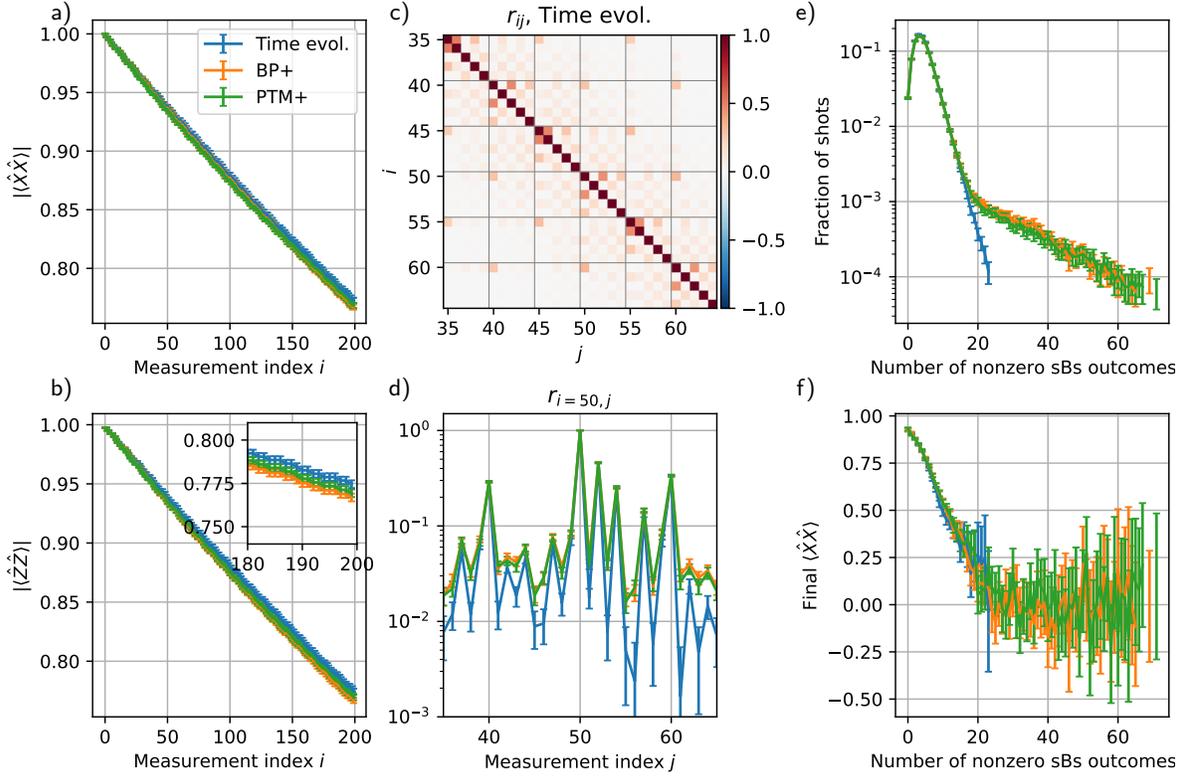

We simulate situations which are computationally accessible with BP+ as well as time evolution simulations, while also being relevant for the surface code.
The largest such simulation is presented here and is shown schematically in fig.~\ref{fig:XX_ZZ_code}, with additional simulations in appendix \ref{app:additional_sims}. 

We compare predictions of the time evolution model of section \ref{sec:noise_model}, and its PTM+ and BP+ approximations, 
for a prototypical two--qubit outer code with stabilizers $\hat X\hat X$ and $\hat Z\hat Z$ which stabilizes a Bell pair.
This qubit code has a unique code state $\ket\psi = \frac{1}{\sqrt{2}} (\ket{00} + \ket{11})$, 
and hence does not encode any logical information.
The code detects a single $\hat X$ and a single $\hat Z$ error acting on either of the two data qubits.
For reasons of computational feasibility, we model one of the two data qubits as a GKP mode, and the other as
an ideal two--level qubit. The CNOT gates between the two--level data qubit and the syndrome qubit are modelled as ideal. 
Each outer--code stabilizer is measured 20 times, 
alternating between measuring $\hat X\hat X$ and $\hat Z\hat Z$. 
After each measurement of the outer--code syndrome qubit, the GKP mode is stabilized for four rounds, 
in the sequence $(\sBs_q, \sBs_p, \sBs_q, \sBs_p)$.
The initial state of the data qubits is set as the code state $\frac{1}{\sqrt{2}} (\ket{00} + \ket{11})$ in the no--error sector.

While the circuit involves few modes and can thus be simulated using all three models, 
it constitutes a meaningful check of the quality of the PTM+ and BP+ approximations in a situation which shares features with the surface code.
Since both $\hat X\hat X$ and $\hat Z\hat Z$ stabilizers are measured, the circuit implementation of the parity check circuits 
involves CNOT gates between the syndrome qubit and data qubits with alternating directions, similarly to surface code parity checks, as shown in fig.~\ref{fig:XX_ZZ_code}.
The circuit is also deep, containing a total of 200 TLS measurements, and thus performs a meaningful check that effects unmodelled by PTM+/BP+ do not grow over the course of deep circuits.

For each of the three models, we perform a Monte--Carlo simulation with $3\cdot 10^5$ shots.
For each shot, the measurement outcomes and associated post--measurement states of the sBs rounds and of the outer--code stabilizer 
measurements are randomly sampled from the appropriate probability distributions.
For the time evolution model, the quantum state for each shot is represented as a pure state, 
and the solution of the Lindblad master equation is sampled from using the quantum trajectories approach~\cite{plenio_quantum-jump_1998}. For the PTM+ and BP+ models, an error--diagonal density matrix is evolved for each 
shot. For every shot, we record the measurement outcomes of the sBs and outer--code stabilizer 
measurements, as well as expectations of the sixteen Pauli operators of the two data qubits after each measurement.
For the GKP qubit, the logical Pauli operator $\hat P_\ell = \hat I_E \otimes \pauli_\ell = \sum_e \pauli_{e\ell}$ determined by the $\sBs$ basis is used for expectation values.

In fig.~\ref{fig:XX_ZZ_results}, various statistics of the resulting data are shown. 
The three models show excellent agreement for the decay of the expectations of the outer--code stabilizers $\hat X \hat X$ and $\hat Z \hat Z$.
From the Pearson correlation coefficients $r_{ij}$ between measurement results 
at rounds $i$ and $j$ we see that such correlations are well modelled by PTM+ and BP+, 
with PTM+ and BP+ somewhat overestimating small correlations. In particular, PTM+ and BP+ successfully model correlations between the inner code syndromes and outer code syndromes, which cannot be modelled by standard Clifford simulations.
PTM+ and BP+ also successfully predict the likelihood of observing a trajectory with a given number of non--zero $\sBs$ outcomes, somewhat overestimating the likelihood of trajectories with a large number of non--zero $\sBs$ outcomes.

An important consideration for concatenated codes is using the 
inner--code syndrome information to help outer code decoding.
To facilitate development of concatenated codes and decoders, 
correlations between inner--code syndromes and logical performance hence have to be modelled accurately.
As a simple measure of these correlations, we analyze the final value of the outer--code stabilizer $\hat X \hat X$, as a function of the number of non--zero 
$\sBs$ outcomes during the trajectory. The three models agree well on this measure.

\section{Surface code simulations}\label{sec:surface_code_sims}
In this section we demonstrate that the BP+ model can be used to estimate the logical error rate 
of a concatenated code. We simulate a rotated surface code of distance $d=5$, 
where the data qubits are GKP qubits, stabilized using an auxiliary TLS, and the syndrome qubits are two--level qubits, 
as depicted in fig.~\ref{fig:surface_code_illustration}.
The surface code circuit is initialized in the logical $\ket{+}$ state of the surface code, the outer--code stabilizer measurements are repeated five times, and all data qubits are measured at the end of the circuit.

We caution that while BP+ could also be used to model preparation, measurement, and idling 
errors, here we have focused on the errors during entangling gates only for simplicity. In addition, the outer--code implementation has not been optimized for performance: a 
more performant implementation may have a different schedule of sBs stabilization rounds, and 
may use a finite--energy measurement of the outer--code stabilizers, as presented in ref.~\cite{royer_encoding_2022}, 
instead of the implementation using CNOTs implemented with a 
single CD gate. In summary, the logical error rates presented here should not be taken as 
representative of the realistic or the best possible performance of a concatenated scheme,
but only as a demonstration of the BP+ method.

\subsection{Sampled error space dynamics}\label{sec:sampled_error_space_dynamics}

\begin{figure}[t]
    \centering
\begin{tikzpicture}
    \draw (0, 0) node[inner sep=0] {\includegraphics[width=1\linewidth]{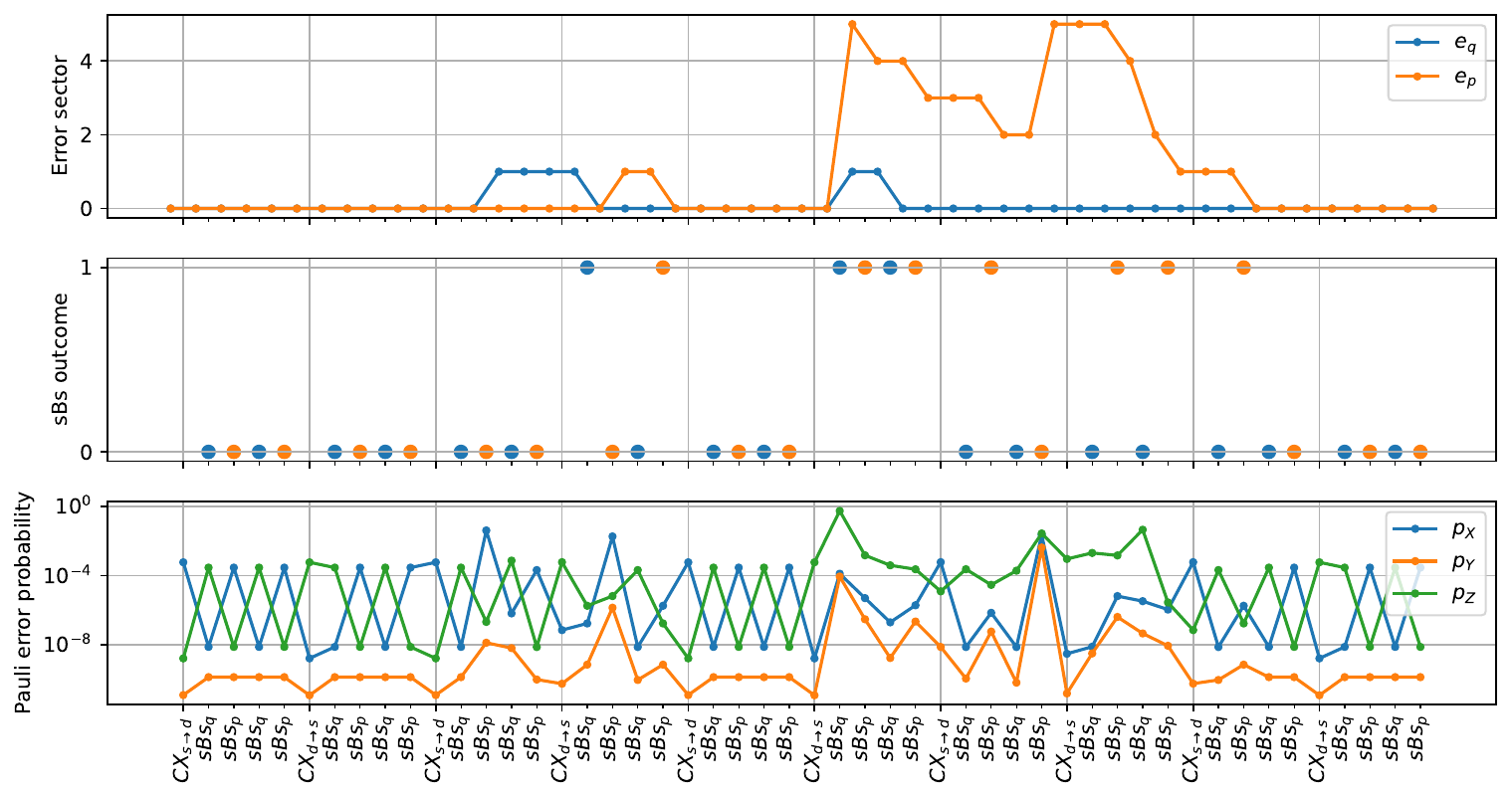}};
    \draw (-7.2, 4.2) node {\small \textsf{a)}};
    \draw (-7.2, 1.5) node {\small \textsf{b)}};
    \draw (-7.2, -0.8) node {\small \textsf{c)}};
 \end{tikzpicture}
    \caption{A manually chosen sampled history of 
    error sector index and $\sBs$ outcomes,
    for one of the data qubits of a $d=5$ rotated surface code.
    a) Sampled error sector index $e = (e_q, e_p)$ of the GKP qubit. Transitions are sampled from the 
    transition matrix $p(o, e|e')$ of the BP+ model of the gate being applied.
    b) sBs outcomes $o$. c) Pauli error probabilities being applied to the logical state, taken from the Pauli error 
    distribution $p(\ell|o, e, e')$ of the gate being applied, conditioned on the input and output error indices shown in panel 
    a), and the sBs outcomes in panel b). 
    The logical errors caused by $CX$ gates are described by 16 
    two--qubit Pauli error probabilities $p_{\ell_d, \ell_s}$, and shown here are the marginal error probabilities $p_{\ell_d} = \sum_{\ell_s}p_{\ell_d, \ell_s}$
    acting on the GKP data qubit.
    }
    \label{fig:error_sector_trajectory}
\end{figure}

Figure \ref{fig:error_sector_trajectory} shows an example of the randomly sampled evolution of the error sector index of one of 
the surface code data qubits over the course of the surface code circuit. 
Also shown are the outcomes of the $\sBs$ stabilization
rounds of this GKP qubit, and the Pauli error probabilities acting on the logical information.
This data is obtained by following the algorithm ~\ref{algo:classical_part}.
The random seed leading to this evolution has been manually chosen to illustrate the features which may occur.

Generally speaking, there is competition between processes which increase and which decrease the error sector index.
Processes which increase the index include physical errors captured in the noise model, 
as well as transitions caused by the approximate nature of the $\sBs$ basis, the $\sBs$ protocol, and the CNOT implementations.
The main mechanism for decreasing the error sector index is the $\sBs$ protocol.

We also observe  {from fig.~\ref{fig:error_sector_trajectory} c)} that $\sBs_p$ and $CX_{s\rightarrow d}$ have larger $\hat X$ error rates than $\hat Z$ error rates.
This may be understood as follows: Both gates feature a large conditional displacement in the $q$ direction, 
and decay of the participating TLS during this conditional displacement can cause a large unwanted displacement in the $q$ direction.
Such a displacement acting on the GKP $\ket{0_\Delta}$ state can cause it to have large overlap with the $\ket{1_\Delta}$ state and vice versa, 
thus possibly causing an $\hat X$ error. 
This effect, which is an important limitation of current GKP code implementations based on conditional displacement gates \cite{lachance-quirion_autonomous_2023, sivak_real-time_2023}, 
is captured by the physical noise model we employ here, 
as well as the BP+ models distilled from it. 
For similar reasons, $\sBs_q$ and $CX_{d\rightarrow s}$ have enhanced $\hat Z$ error rates.

We also see that a decrease in error sector index is often, but not always, associated with 
a non--zero sBs measurement outcome in the same quadrature, 
and that generally, error rates are higher if the ``input" or 
``output" error sector is nontrivial.

\subsection{Logical outer--code error rates}
\begin{figure}
\centering
\includegraphics[width=.7\textwidth]{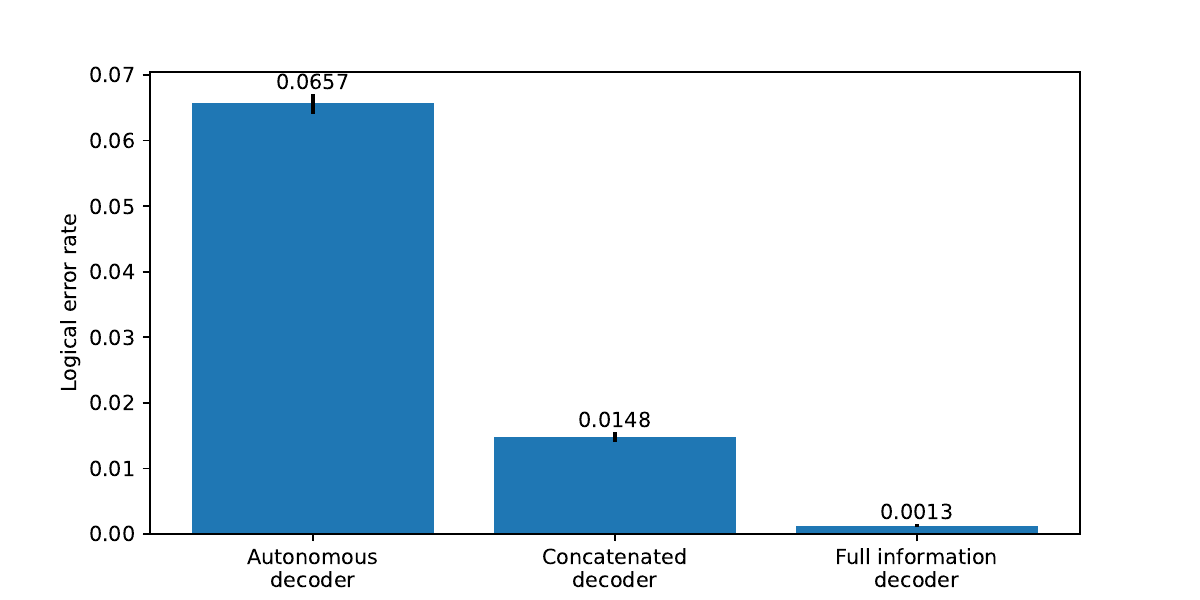}
\caption{As a demonstration of the Bosonic Pauli+ simulation method, logical error rates are obtained for an implementation of the concatenation of a single--mode GKP code stabilized by the $\sBs$ protocol, with a $d=5$ rotated surface code. Physical error rates are given in table \ref{tab:noise_rates}. The same dataset is decoded using three different outer--code decoders: The autonomous decoder discards the inner--code syndrome data returned by the $\sBs$ protocol. The concatenated decoder uses that data to boost outer--code decoding, and the unrealistic ``full information decoder'' uses knowledge of the dynamics on the non--computational part of the GKP Hilbert space which is not accessible in realistic experiments. The error rates should be taken only as a demonstration of the BP+ method, and not as the realistic or the best possible performance of a concatenated GKP code.
}
\label{fig:logical_error_rates}
\end{figure}

In fig.~\ref{fig:logical_error_rates} we present the logical error rates obtained when simulating a $d=5$ rotated surface code quantum memory experiment, with five repetitions of the 
outer--code stabilizer measurements, using BP+. 
We take $10^5$ samples of the $\sBs$ outcomes, outer--code stabilizer measurements, and final data qubit measurements 
using the two--step sampling procedure outlined in section \ref{sec:circ_sim}, and using the BP+ models presented in section \ref{sec:gates}.
For each sample, the $\sBs$ outcomes and outer--code syndrome outcomes are processed by one of three different outer--code decoders. The outer--code decoder suggests a correction operation to apply to the outcomes of the final data qubit measurements. For each decoder, we report the logical error rate, which is the fraction of samples for which the corrected final measurements indicate a surface code logical state which is different from the known initial state.

All of the three decoders are minimum--weight perfect matching decoders, and 
build on the \emph{stim} and \emph{pymatching} software packages \cite{gidney2021stim, higgott_sparse_2023}.
The ``autonomous decoder'' does not use the outcomes of $\sBs$ rounds, and is 
applicable even when $\sBs$ outcomes are not experimentally accessible \cite{lachance-quirion_autonomous_2023}.
The ``concatenated decoder" uses knowledge of the $\sBs$ outcomes, which allows more accurate outer--code decoding and better outer--code performance compared to the autonomous decoder,
as shown in fig.~\ref{fig:logical_error_rates}. That inner--code syndrome information can boost outer code performance was also observed for a different inner--code GKP error correction protocol in 
ref.~\cite{lin_closest_2023} and earlier work.

Lastly, the ``full information'' decoder assumes knowledge about the error sector index of 
every mode before and after every gate of the outer--code circuit, in addition to the sBs 
and outer--code syndrome outcomes.
It is implemented by decoding over the sampling circuit returned by algorithm \ref{algo:classical_part}.
It can be understood as inserting perfect ``which--error--sector"
measurements before and after each gate, and providing their outcomes to the outer--code decoder.
While this decoder is not realistic, since ``which--error--sector" information is not experimentally 
accessible, it is interesting to quantify the improvement such information would lead to.

\section{Conclusion}\label{sec:conclusion}
We have introduced Bosonic Pauli+ (BP+), a model and simulation method tailored to simulating GKP qubits stabilized using the $\sBs$ protocol, 
and concatenated with a qubit outer code.
BP+ models are extracted from time--evolution simulations using realistic physical noise models. BP+ does not require the use of simplified analytic noise models such as Gaussian displacement noise, specifying the Krauss operators of noise channels, or approximating the GKP qubits as two--level logical systems.
We demonstrated the end--to--end use of the BP+ method to obtain logical error rates of a concatenated code implementation,
for a particular implementation of the concatenated codes and a given physical noise model.
We reiterate that the reported logical error rates should be taken as a demonstration of the BP+ method, and
not as predictions about the realistic or the best possible performance of a concatenated GKP code.

We expect that BP+ will be useful to study the impact of physical decoherence rates and implementation choices on the logical error rate of a concatenated GKP code.
Implementation choices which can be studied include implementation details of the $\sBs$ stabilization or other stabilization protocols,
how many rounds of the $\sBs$ protocol to run for each step of the outer code error correction,
the choice of outer qubit code and outer code decoder, and the implementation of CNOT gates. We also expect that describing the dynamics of the non--logical degrees of freedom of a GKP mode as a classical random process can be useful conceptually, even in the absence of an outer code.

We have also introduced the $\sBs$ basis, a Hilbert space basis for the finite--energy GKP code which gives a clean description of the quantum dynamics caused by $\sBs$ stabilization.
We expect that the $\sBs$ basis will be useful also outside the context of simulating concatenated codes, 
for example for visualizing and optimizing gates and operations for GKP codes at the quantum control level.

Several extensions of the BP+ method are possible.
Within the realm of GKP qubits, it may be possible to study gates between two GKP modes, 
which would be necessary for an ``all--GKP'' outer code which uses GKP qubits both as data and syndrome qubits.
It may also be possible to model finite--energy multi--qubit Pauli measurements as presented in ref.~\cite{hastrup_improved_2021} using BP+, 
as a replacement for the outer--code parity check circuit simulated here, or to model some 
multi--mode GKP codes \cite{royer_encoding_2022, conrad_gottesman-kitaev-preskill_2022}.
Because of the larger involved Hilbert space sizes and number of modes, these extensions would require solving computational challenges, 
in order to extract the BP+ parameters from a more detailed model and in order to sample from the resulting BP+ channels in a tractable way.

Extensions may also apply BP+ to concatenations of bosonic codes other than the GKP code,
such as dissipative cat codes \cite{mirrahimi_dynamically_2014}, Kerr--Cat codes \cite{cochrane_macroscopically_1999, puri_engineering_2017}, binomial codes \cite{michael_new_2016} and others \cite{albert_bosonic_2022}. 
The main challenge for these cases will be finding a basis to
replace the $\sBs$ basis, such that the dynamics on the
error subspace can be approximately understood in terms of classical population transfer rates, 
and do not involve coherent quantum effects.

\subsection*{Data availability and Acknowledgements}
 {Data underlying the figures is available from the authors on reasonable request.}
The authors thank Guillaume Duclos--Cianci for helpful comments.
B.R. acknowledges support from NSERC,
Fonds de recherche du Québec Nature et Technologies,
and the Army Research Office under Grant W911NF2310045.
This research was undertaken thanks in part to 
funding from the Canada First Research Excellence Fund.

\bibliographystyle{quantum}
\bibliography{references}

\newpage 

\appendix

\section{From PTM+ to BP+}\label{app:ptmp_to_bpp}

A BP+ channel is a PTM+ channel, 
whose logical action for each error sector transition
is further approximated as a Pauli error channel, 
$
    \hat \rho 
    \mapsto 
    \sum_\ell p_\ell \pauli_\ell \hat \rho \pauli_\ell
$.
We define a BP+ channel as the Pauli twirling of a PTM+ channel:
\begin{equation}
    \bpp (\hat{\rho})
    = 
    \frac{1}{4^N} 
    \sum_\ell \hat P_{\ell}
    \ptmp (\hat P_\ell \hat{\rho} \hat P_\ell) 
    P_{\ell}^\dag
\end{equation}
where $\hat P_\ell = \sum_e \pauli_{e\ell} = \hat I_E \otimes \pauli_{\ell}$ are the logical Pauli operators.

In this appendix, we show that twirling indeed maps PTM+ to BP+, and 
compute the BP+ coefficients of eq.~\eqref{eq:bpplus_action_on_density_matrix} from the PTM+ coefficients of eq.~\eqref{eq:ptmplus_action}.
We refer the reader to refs.~\cite{bennett_mixed-state_1996, geller_efficient_2013, wood_tensor_2015} 
for background and proofs on Pauli twirling and the different channel representations used here.

To compute the twirling,
we first convert the Pauli transfer matrix representation of a PTM+ channel
into the process matrix representation. The action of the channel in the process matrix representation is
\begin{align}
    \ptmp(\hat \rho) 
    = \sum_{e e' \ell \ell'} 
    \chi_{e e' \ell \ell'} 
    \pauli_{ee' \ell}
    \hat{\rho} 
    \pauli_{e e' \ell'}^{\dag}.
\end{align}
Here $\pauli_{e e' \ell} = \op{e}{e'} \otimes \pauli_\ell$ combines a Pauli operation and an error sector transition.
The $S$ and $\chi$ tensors are related by eq.~\eqref{eq:coeff_ptm_trace} as
\begin{equation}
    S_{e e' \ell \ell'}  
    = \frac{1}{2^N}
    \sum_{m m'} \chi_{e e' m m'} T_{\ell \ell' m m'}
\end{equation}
with $
    T_{\ell \ell' m m'}
    =
    \tr\qty(
        \pauli_\ell \pauli_m \pauli_{\ell'} \pauli_{m'}
    ).
$
Viewing $T_{[\ell \ell'][m m']}$ as a matrix,
we invert the above relation to obtain $\chi$ from $S$ as
\begin{equation}
    \chi_{ee'\ell \ell'}
    =
    2^N 
    \sum_{mm'} S_{ee'\ell \ell'} T^{-1}_{\ell \ell' m m'}.
\end{equation}

Pauli twirling zeroes the off--diagonal elements 
$\ell \neq \ell'$ of $\chi$, for each error space transition $e'\rightarrow e$.
This gives a channel of the form
\begin{equation}
    \bpp(\hat{\rho})
    =
    \sum_{ee'\ell} \chi_{ee'\ell} 
    \pauli_{ee'\ell} \hat{\rho} \pauli_{ee' \ell}^\dag
\end{equation}
with $\chi_{ee'\ell} = \chi_{ee'\ell \ell}$.

Finally, we set $p(e | e') = \sum_\ell \chi_{ee'\ell}$, and $p(\ell|e, e') = \chi_{ee'\ell} / p(e|e')$ to arrive at the parametrization of 
eq.~\eqref{eq:bpplus_action_on_density_matrix}.
The normalization conditions $\sum_e p(e|e') = \sum_\ell p(\ell|e, e') = 1$ follow from the fact that the twirled channel preserves the trace.

\section{Gauge}\label{app:gauge}
The implementation of CNOT gates described in section \ref{sec:cnot} has the additional effect of displacing the peaks of the GKP state in a way that leaves the GKP code space.
This appendix explains the generalizations of the algorithms and equations in the main text which are required to account for this effect.

The effect can be understood by observing that the implementation eq.~\eqref{eq:ecd_model} of the gate $CD(\beta)$ is not trivial when the TLS is in its 
ground state: rather the bosonic mode is displaced by $-\beta/2$. This is in contrast to the desired CNOT gate, whose action should be trivial when the TLS is in its ground state.

While this effect could in principle be corrected by an additional unconditional displacement operation, 
doing so would leave the finite--energy envelope of the output state off--center, likely resulting in sub--optimal performance.
Instead, here we adapt the concept of gauge, from the setting of multi--mode GKP codes as presented in refs.~\cite{royer_encoding_2022, conrad_gottesman-kitaev-preskill_2022}, 
to our setting of single--mode GKP codes.

The gauge degree of freedom can be described as a tuple $(g_q, g_p)$, with $g_q, g_p \in \{0, 1\}$. 
These variables indicate that the code states are simultaneous eigenstates of $\hat S^q_\Delta$ and $\hat S^p_\Delta$, 
with eigenvalues $e^{i \pi g_q}$ and $e^{i \pi g_p}$ respectively.
The analytical form of the ``gauged'' finite--energy states are given by the following generalization of eq.~\eqref{eq:analytic_fe_states}:
\begin{align}
    \ket{\mu_{\Delta, g_q, g_p}} = \frac{1}{\mathcal N_{\mu, \Delta, g_q, g_p}} e^{-\Delta \hat n} \sum_{k=-\infty}^\infty e^{-i l \cdot \tfrac14 g_q \hat x} \ket{\hat 
    x = (k + \tfrac12 \mu + \tfrac14 g_p l},
\end{align}
where $l$ is the GKP lattice spacing.

When acting on a code space with non--zero gauge, the $\sBs$ protocols need to be modified to perform the correct stabilization: 
When $g_q = 1$, the protocol for $\sBs_q$ in fig.~\ref{fig:sbs_circuit} is modified by flipping the sign of the third conditional 
displacement, and replacing the last gate on the TLS by $\hat R_Y(+\pi/2)$.
When $g_p = 1$, the protocol for $\sBs_p$ is updated in the same way.

Likewise, the implementation of the CNOT gates requires modification for a non--zero input gauge: 
Defining the signs $s_{q, p} = 2 g_{q, p} - 1$, and the Pauli operator $\hat Z$ of the TLS, the implementations of eq.~\eqref{eq:cx_implementations} can be generalized to
\begin{align}
    CX_{s\rightarrow d} ={}& e^{-i \frac \pi 4 \cdot g_p s_q \hat Z} \circ \widetilde{CD} ( s_q \cdot \sqrt{\pi}) \\
    CX_{d\rightarrow s} ={}& H_s \circ e^{i \frac \pi 4 \cdot s_p g_q \hat Z}  \circ \widetilde{CD} ( s_p \cdot -i \sqrt{\pi} ) \circ H_s.
\end{align}
After a CNOT gate, the state will have different gauge compared to the input state: $CX_{s\rightarrow d}$ sends $g_q \rightarrow 1-g_q$, and $CX_{d\rightarrow s}$ sends $g_p \rightarrow 1-g_p$.

Using the gauged finite--energy states and the Kraus operators of the gauged sBs protocols, an $\sBs$ basis can be built for each of the four possible gauges following algorithm \ref{algo:sBs_basis_algo},
leading to a gauged generalization $\pauli_{e \ell}^{g_q, g_p}$ of the operators defined in eq.~\eqref{eq:pauli_error_basis}. Taking into account that a channel $\channel$ 
may have a different input and output gauge $g^\text{in}$ and $g^\text{out}$, the PTM+ matrix elements may then be extracted by generalizing eq.~\eqref{eq:coeff_ptm_trace}:
\begin{equation}
    S_{ee'\ell\ell'} = \frac{1}{2^N} \tr \left(\pauli_{e\ell}^{g_q^\text{out}, g_p^\text{out}} \channel (\pauli_{e'\ell'}^{g_q^\text{in}, g_p^\text{in}}) \right).
\end{equation}

For the model comparison simulations of section \ref{sec:gaining_confidence}, 
when computing the time evolution with the detailed model, the gauge is kept track of as outlined here.
For the PTM+ and BP+ simulations, only a single PTM+/BP+ is computed for every class of gate, with the trivial input gauge,
and is used in the simulation regardless of what the gauge actually is.
The underlying assumption is that the PTM+ and BP+ models with different input gauges are very similar. The fact that results match well between the three models supports this assumption.

\section{Additional model comparison}\label{app:additional_sims}

\begin{figure}[t]
\centering
\begin{tikzpicture}
    \draw (0, 0) node[inner sep=0] {\includegraphics[width=\textwidth]{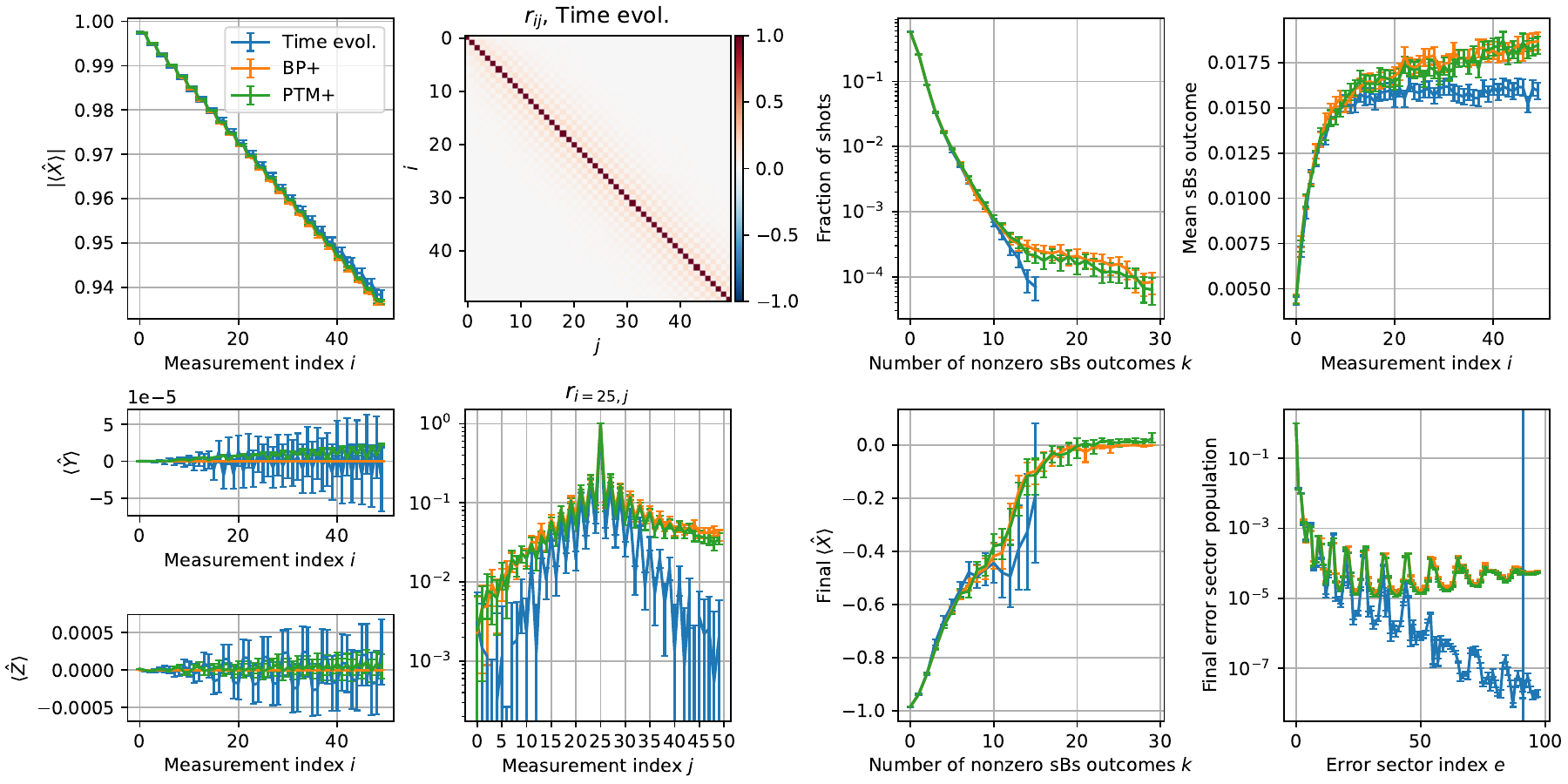}};
    \draw (-7.3, 4) node {\small \textsf{a)}};
    \draw (-7.3, 0) node {\small \textsf{b)}};
    \draw (-3.5, 4) node {\small \textsf{c)}};
    \draw (-3.5, 0) node {\small \textsf{d)}};
    \draw (0.9, 4) node {\small \textsf{e)}};
    \draw (0.9, 0) node {\small \textsf{f)}};
    \draw (4.8,4) node {\small \textsf{g)}};
    \draw (4.8,0) node {\small \textsf{h)}};
 \end{tikzpicture}
\caption{Simulation data from simulating 50 rounds of stabilization of a single GKP mode, alternating between $\sBs_q$ and $\sBs_p$.
The initial state is $\ket{+X}$ in the no--error sector.
a) Absolute expectation of $\hat X$ after each measurement, for each of the three models.
b) Expectations of $\hat Y$ and $\hat Z$ after each measurement.
c) Heatmap of Pearson correlation coefficients $r_{ij}$ between $\sBs$ outcomes at measurements indexed $i$ and $j$, for the time evolution model.
d) Slice through c) at $i=25$, for the three different models.
e) Fraction of shots with exactly $k$ non--zero $\sBs$ outcomes
f) Post--selected value $\langle \hat X \rangle$ after the final measurement, as a function of the number of non--zero $\sBs$ outcomes $k$.
g) Mean values of $\sBs$ measurement outcomes. 
h) Mean error sector populations for the state after the final measurement.
The error sectors whose corresponding $\sBs$ basis vectors have been drawn randomly in algorithm \ref{algo:sBs_basis_algo} are to the right of the vertical line.
}
\label{fig:stab_model_match_data}
\end{figure}
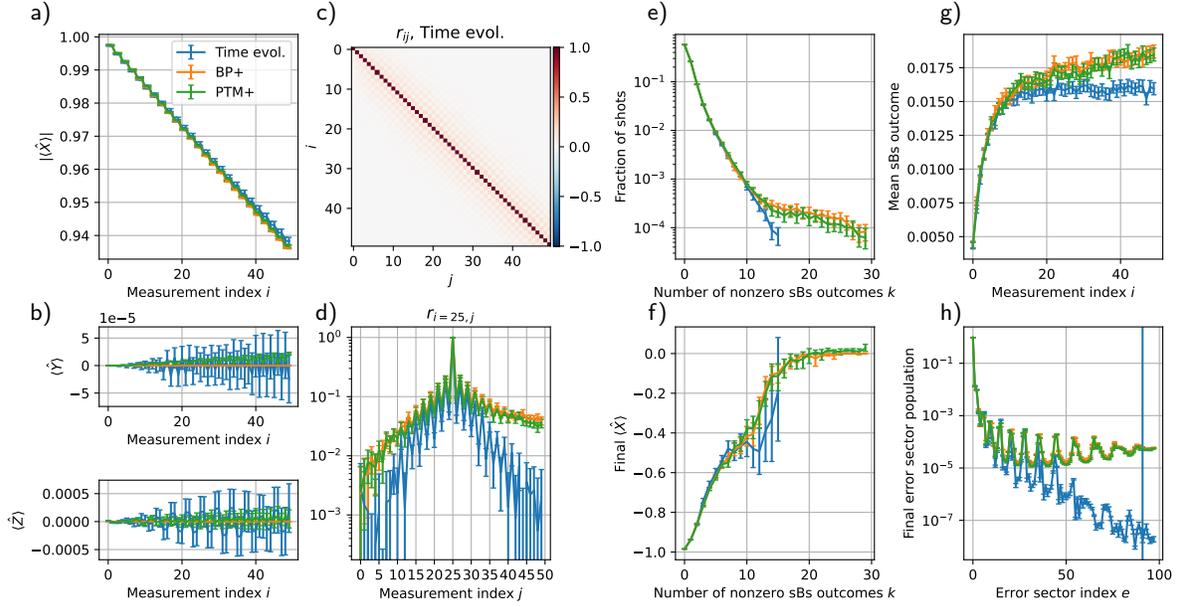

\begin{figure}[t]
\centering
\begin{tikzpicture}
    \draw (0, 0) node[inner sep=0] {\includegraphics[width=.8\textwidth]{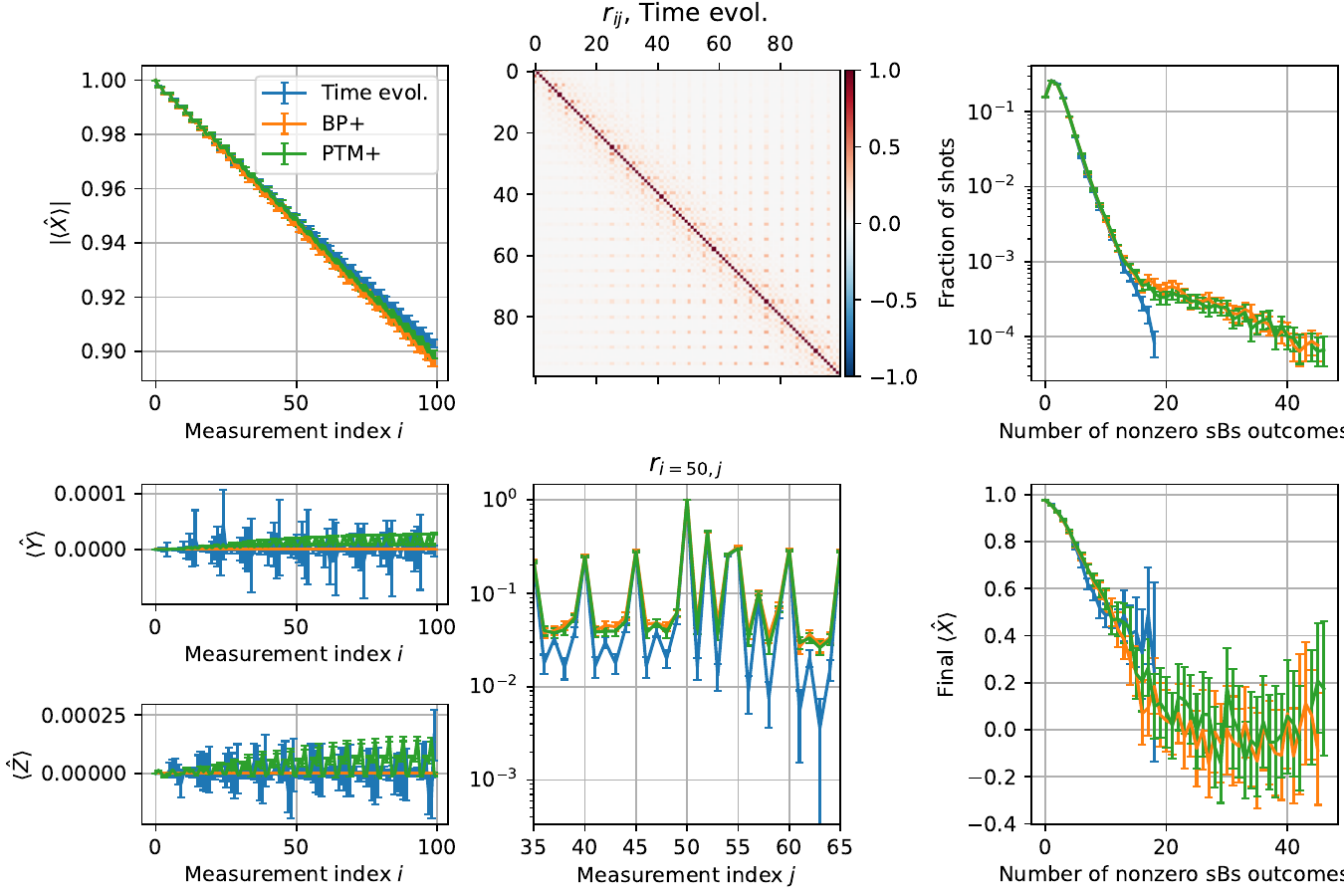}};
    \draw (-5.5, 3.8) node {\small \textsf{a)}};
    \draw (-5.5, 0) node {\small \textsf{b)}};
    \draw (-1.7, 3.8) node {\small \textsf{c)}};
    \draw (-1.7, 0) node {\small \textsf{d)}};
    \draw (2.7, 3.8) node {\small \textsf{e)}};
    \draw (2.7, 0) node {\small \textsf{f)}};
 \end{tikzpicture}
 \caption{Simulation data from simulating 25 repetitions of measuring the GKP operator $\hat X$ using an auxiliary TLS, using the three different models. After each measurement, four rounds of $\sBs$ are simulated in the sequence $(\sBs_q, \sBs_p, \sBs_q, \sBs_p)$.
The initial state is $\ket{+X}$ in the no--error sector. Measurement indices $5 \cdot k, k \in \mathbb N_0$ correspond to the logical measurement, and the remaining round indices correspond to $\sBs$ rounds.
a) Absolute expectation of $\hat X$ after each measurement, for the three different models. 
b) Expectations of $\hat Y$ and $\hat Z$ after each measurement.
c) Heatmap of Pearson correlation coefficients between outcomes at different rounds, for the time evolution model.
d) Slice through c) at i=25, for the three different models.
e) Proportion of trajectories with exactly $k$ non--zero $\sBs$ outcomes, as a function of $k$.
f) Post--selected value $\langle \hat X \rangle$ after the final measurement, as a function of the number of non--zero $\sBs$ outcomes.}
\label{fig:repeated_logical_meas_model_match}
\end{figure}

We provide two further examples of Monte--Carlo simulations of deep circuits, comparing the predictions of time evolution simulations with BP+ and PTM+.
For both circuits, as in the main text, each measurement outcome is randomly sampled according to the outcome probabilities predicted by the model, and we take $3\cdot10^5$ Monte--Carlo 
shots for each model and each initial state.
The measurement outcomes and expectation values of logical operators are recorded after each round. Each shot of the open--system time evolution simulations is simulated using a quantum trajectories solver,
while for the PTM+ and BP+ models, error--diagonal density matrices are evolved for each shot.

The first circuit we consider acts on a single bosonic mode, and contains 50 rounds of stabilization, alternating between $\sBs_q$ and $\sBs_p$.
The starting state is set to be one of the six cardinal states of the GKP qubit, in the no--error sector $e=0$. Figure \ref{fig:stab_model_match_data} shows results for the $\ket{+X}$ initial state, and results for other initial states are qualitatively similar.

There is good agreement between all three models for the decay of logical information $\langle \hat X \rangle$, with the approximate models PTM+ and BP+ being slightly too pessimistic.
Predictions for $\langle \hat Y \rangle$ and $\langle \hat Z \rangle$ are mostly within error bars of each other, and remain small. Note that BP+ cannot model the non--unital effects by which an initially zero expectation value could become non--zero.
PTM+ and BP+ predict the larger correlations between sBs outcomes at different rounds well, and in particular capture that same--quadrature rounds are more strongly correlated than opposite--quadrature rounds. The approximate models overestimate the small correlations, as well as the time scale at which correlations decay. The correlation data is in qualitative agreement with the experimental data of ref.~\cite{sivak_real-time_2023}, appendix S4e.
The approximate models also overestimate the average probability of finding a non--zero sBs outcome in a given round, overestimate the probability of trajectories having a large number $k$ of non--zero sBs outcomes, and overestimate the populations of higher error sectors after the final measurement.
The probabilities of trajectories having a small number $k$ of non--zero sBs outcomes, and populations of the lower error sectors, are modelled accurately,
as is the relationship between the number of non--zero sBs outcomes in a trajectory, and the final logical expectation.
Overall, for this circuit the approximate models PTM+ and BP+ are too pessimistic, compared to time evolution simulations, and results of PTM+ and BP+ are almost indistinguishable.
For this circuit, we have also checked in a separate experiment that decreasing the Fock cutoff and number of ranks in the sBs basis does not importantly influence the predictions.

The second circuit also acts on a bosonic mode starting from the no--error $\ket{+X}$ state, and simulates 25 rounds of logical measurement in the $X$--basis, which is implemented using a TLS ancilla and the $CX_{s\rightarrow d}$ gate. Each logical measurement is followed by four rounds of stabilization. Results are shown in fig.~\ref{fig:repeated_logical_meas_model_match}.
Again, the models show good agreement for the decay of logical information $\langle \hat X \rangle$, with the approximate models PTM+ and BP+ being slightly too pessimistic.
Predictions for the evolution of $\langle \hat Y \rangle$ and $\langle \hat Z \rangle$ disagree between models, but remain small and are unlikely to affect the functioning of an outer code.
Again, PTM+ and BP+ capture larger correlations well and overestimate smaller correlations.
PTM+ and BP+ overestimate the likelihood of trajectories with many non--zero sBs outcomes, and the three models mostly agree on the relationship between the number of non--zero sBs outcomes and the final logical expectation.

\end{document}